\documentclass[iop]{emulateapj}
\usepackage{natbib}
\usepackage{float}
\begin{document}


\title{Chemical Abundances in NGC~5053: A Very Metal-Poor and Dynamically Complex Globular Cluster}

\author{Owen M. Boberg\altaffilmark{1}, Eileen D. Friel\altaffilmark{1}, Enrico Vesperini }
\affil{Astronomy Department, Indiana University,
    Bloomington, IN 47405}

\altaffiltext{1}{Visiting Astronomer, Kitt Peak National Observatory, 
National Optical Astronomy Observatory, which is operated by the Association 
of Universities for Research in Astronomy, Inc. (AURA) under cooperative
agreement with the National Science Foundation.}    

\begin{abstract}

NGC~5053 provides a rich environment to test our understanding of the 
complex evolution of globular clusters (GCs).
Recent studies have found that this cluster has interesting 
morphological features beyond the typical spherical distribution of 
GCs, suggesting that external tidal effects have played an important 
role in its evolution and current properties.
Additionally, simulations have shown that NGC~5053 could 
be a likely candidate to belong to the Sagittarius dwarf galaxy (Sgr dSph) stream. Using the 
Wisconsin-Indiana-Yale-NOAO-Hydra multi-object spectrograph, we have collected 
high quality (signal-to-noise ratio $\sim$ 75-90), medium-resolution 
spectra for red giant branch stars in NGC~5053. Using these spectra we 
have measured the Fe, Ca, Ti, Ni, Ba, Na, and O abundances in the cluster. We 
measure an average cluster [Fe/H] abundance of -2.45 with a standard 
deviation of 0.04 dex, making NGC~5053 one of the most 
metal-poor GCs in the Milky Way (MW). The [Ca/Fe], [Ti/Fe], and [Ba/Fe] we measure 
are consistent with the abundances of MW halo stars at a 
similar metallicity, with alpha-enhanced ratios and slightly depleted [Ba/Fe]. 
The Na and O abundances show the Na--O anti-correlation found in most GCs. From 
our abundance analysis it appears that NGC~5053 is at least chemically 
similar to other GCs found in the MW. This does not, however, rule out 
NGC~5053 being associated with the Sgr dSph stream.

\end{abstract}


\section{Introduction}
\begin{deluxetable*}{ccccccccccc}[H]
\tabletypesize{\scriptsize}
\tablecaption{Comparison of NGC~5053 and NGC 5024 (M53) Properties\label{tbl-1}}
\tablewidth{\textwidth}
\tablehead{\colhead{Cluster} &\colhead{R.A.}& \colhead{Decl.} & \colhead{($l,b$)} & \colhead{$R_{\odot}^{a}$} & \colhead{$R_{\mathrm{GC}}^{a}$} 
& \colhead{$r_{c}^{b}$} & \colhead{$r_{h}^{b}$} & \colhead{$r_{\mathrm{tr}}^{b}$} & \colhead{c}  \\
\colhead{} &\colhead{(J2000)}& \colhead{(J2000)} & \colhead{} & \colhead{(kpc)} & \colhead{(kpc)} 
& \colhead{(pc)} & \colhead{(pc)} & \colhead{(pc)} & \colhead{$\mathrm{log}(\frac{r_{tr}}{r_{c}})$}}
\startdata
NGC~5053  & 13:16:27.09 & +17:42:00.9 & $336^{\circ}$,$79^{\circ}$ & 17.4 & 17.8 &  9.71 & 12.39  & 89.13 & 0.96 \\
NGC 5024  & 13:12:55.24 & +18:10:05.4 & $333^{\circ}$,$80^{\circ}$ & 17.9 & 18.4  & 2.18 &  5.84  & 239.9 & 2.04 
\enddata
\tablecomments{$^{a}$Harris (1996, Version 2010),$^{b}${\cite{2005ApJS..161..304M}}}
\end{deluxetable*}

Globular clusters (GCs) have long served 
as our laboratories to study the dynamics 
and evolution of simple stellar populations. 
With the greater availability of high quality and 
homogeneous photometric and spectroscopic data, 
it has become evident that GCs are increasingly
complex stellar populations. The Galactic GC population has shown
that most GCs exhibit abundance patterns and color-magnitude diagram (CMD) morphology
indicative of multiple populations 
\citep[see e.g.,][]{2012A&ARv..20...50G, 2014arXiv1410.4564P}. 
A number of possible sources for the gas out of which 
second generation (SG) stars might have formed have been suggested. These sources 
include rapidly rotating massive stars, massive binary stars, and intermediate-mass 
AGB stars \citep[see e.g.,][]{2001ApJ...550L..65V,2006A&A...458..135P,2008MNRAS.391..825D, 
2010MNRAS.407..854D,2012MNRAS.423.1521D,2009A&A...507L...1D}
and all the aspects of the formation, chemical, and dynamical evolution are 
currently a matter of intensive investigation.

As GCs orbit the Galaxy they are stripped of their stars and 
help populate the halo. In some instances, 
wide-field photometric surveys, such as the
Sloan Digital Sky Survey (SDSS), have revealed 
large tidal tails extending from GCs, giving us
a snapshot of the tidal stripping experienced by GCs. 
The most famous of these examples is Palomar 5 \citep{2001ApJ...548L.165O}. 
Further exploration of SDSS data has revealed that other clusters 
also show large tidal features. 
NGC~5053 is one of these clusters
\citep[see][]{2006ApJ...651L..33L,2010A&A...522A..71J}.

NGC~5053 is a metal-poor GC located near the 
north Galactic cap ($l = 336^{\circ} , b=79^{\circ}$) with a Galactocentric radius
of $R_{GC} = 17.8$ kpc. Near NGC~5053, $\sim$ 500 pc away based on the X,Y and Z positions 
from Harris (1996, 2010 version) is NGC 5024 (M53), located 
at ($l = 333^{\circ} , b=80^{\circ}$) with a $R_{GC} = 18.4$ kpc. 
A comparison of the physical characteristics 
of these clusters is given in Table 1. In Figure 1, a history of the previously 
determined [Fe/H] abundances of NGC~5053 are plotted versus the year of publication. 
At one point NGC~5053 was considered the most metal-poor GC in the Milky Way (MW) 
\citep{1985ApJ...293..424Z}.

Using SDSS photometry, \cite{2006ApJ...651L..33L} discovered 
a $6^{\circ}$ tidal tail associated with NGC~5053. 
A wider field analysis of SDSS photometry,
by \cite{2010A&A...522A..71J}, confirmed the findings 
and further mapped the extra-tidal features of the cluster. 
A study by \cite{2010AJ....139..606C},
using  Megacam on the Canada-France-Hawaii Telescope, 
detected a tidal bridge-like structure between NGC~5053 and its neighbor M53. If this
bridge-like structure is genuine, it suggests that the 
evolution of NGC~5053 has been influenced not only by the Galaxy, but also 
through interactions with M53. It 
should be noted that \cite{2010A&A...522A..71J} do not detect 
this bridge-like structure. The possible complex dynamical history of
NGC~5053 does not stop there. Using simulations of the tidal 
disruption of the Sagittarius dwarf galaxy (Sgr dSph), 
\cite{2010ApJ...718.1128L} identify
NGC~5053 as a likely candidate GC that belongs to the Sgr dSph.

\begin{figure}[H]
\epsscale{1.0}
\includegraphics*[width=0.5 \textwidth]{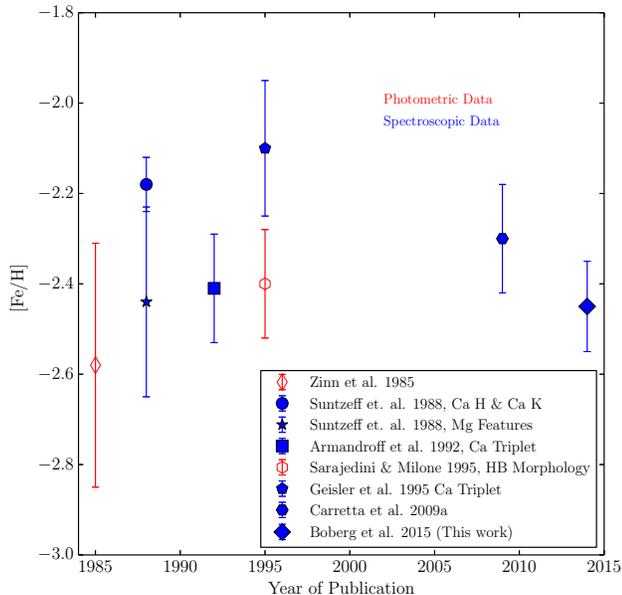}
\caption{Previous [Fe/H] determinations for NGC~5053 taken from the literature along with our results. 
Abundances determined from photometric data are represented by open
red markers, and those determined by spectroscopic data are represented by 
solid blue markers.\label{fig1}}
\end{figure}

Motivated by the apparently complex dynamics of NGC~5053, and its 
possible association with the Sgr dSph,
we have collected high quality spectra of red giant branch (RGB) stars 
in the cluster. With these spectra we have been able to directly measure 
the Fe, Ni, Ca, Ti, Na, and O abundances 
of clusters members in order to determine if NGC~5053
exhibits the typical abundance patterns shown in GCs, 
such as the Na--O anti-correlation. We will compare these abundances with those seen in 
other Galactic GCs and field stars at a similar metallicity. 
We will also present
how the abundances in NGC~5053 compare to the abundances in GCs that 
are generally considered part of the
Sgr dSph \citep[M54, Arp~2, Terzan~7, and Terzan~8; see e.g][]{2014A&A...561A..87C}.

This paper is organized in the following manner. In Section 2 we will 
present how our observations and
data reductions were performed, followed by a description of how 
cluster members were chosen for this study.
In Section 3 we describe how the abundance analysis was carried out 
and how we assessed
the errors in our measurements. The results of our analysis are 
presented in Section 4.
In Section 5 we compare our results for NGC~5053 with abundances in 
other MW and Sgr GCs and fields stars. We will conclude with possible 
implications of these abundances and future studies to come.

\section{Observations and Data Reduction}

\subsection{Observations}
The spectroscopic data for this study were collected 
using the Wisconsin-Indiana-Yale-NOAO (WIYN) 3.5 m
telescope\footnote{The WIYN Observatory is a joint 
facility of the University of Wisconsin-Madison, 
Indiana University, the National Optical 
Astronomy Observatory and the University of Missouri.} 
with the Hydra multi-object spectrograph 
and a 2600 $\times$ 4000 pixel CCD (STA 1) 
on 2014 January 18 and 2014 February 14. The Bench Spectrograph
was used with the ``316@63.4'' grating, resulting 
in spectra with a dispersion of 0.16 \AA $\, \mathrm{pixel^{-1}}$ covering a wavelength
range of approximately $6063 - 6380$ \AA. 
The typical signal-to-noise (S/N) ratios of the spectra were 70-90 for the majority 
of the stars in the sample. In instances where 
a star was observed in multiple Hydra configurations, once on each of the nights, 
S/N ratios are on the order of 100. In addition to our program objects,
high S/N radial velocity standards were observed at the beginning of
each night along with a hot, rapidly rotating 
B star to use as a template for removing telluric features. 

The data were reduced using the standard 
IRAF\footnote{IRAF is distributed by the National Optical Astronomy Observatories,
which are operated by the Association of Universities for 
Research in Astronomy, Inc., under cooperative agreement with the National Science Foundation.} 
software to perform bias subtraction, flat fielding, 
dark subtraction, and dispersion correction with ThAr 
comparison lamp spectra. The full integration 
time (4.5 hr per configuration) was broken up into six separate exposures to reduce the 
effects of cosmic rays in the final combined spectra. 
The processed spectra from each exposure were combined 
using the IRAF task \textit{scombine}.

\subsection{Target Selection}

\begin{figure}
\epsscale{1.0}
\includegraphics*[width=0.5 \textwidth]{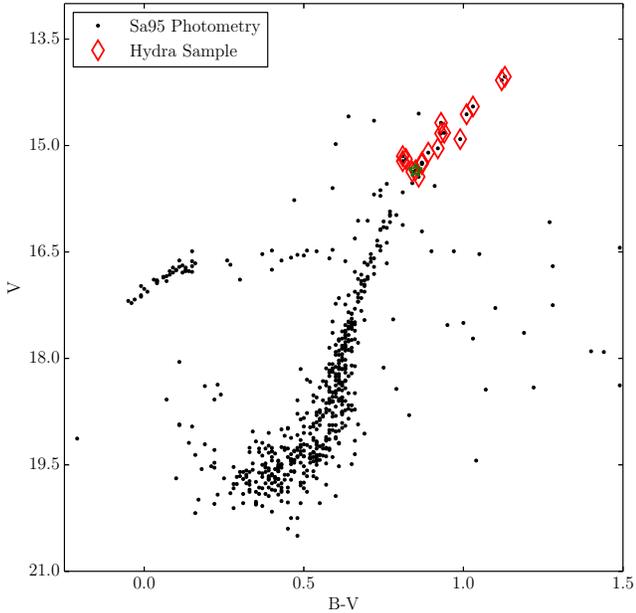}
\caption{$V$ vs $(B-V)$ CMD for NGC~5053. Data taken from \cite{1995AJ....109..269S}. 
The points overplotted with open diamonds are those
stars for which we collected spectra. These are the stars that met the photometric 
criteria described in the paper. The point plotted
as a star marks the location of a star that photometrically appeared to be part 
of the cluster, but its radial velocity ruled it
out as a member\label{fig2}.}
\end{figure}

\begin{figure}
\epsscale{1.0}
\plotone{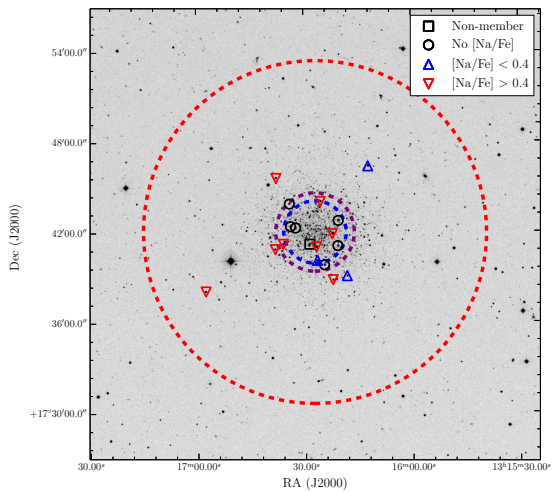}
\caption{A Digitized Sky Survey (DSS) image showing NGC~5053. 
The targets for this study are highlighted by small circles, triangles, and
a square. The triangles mark stars for which we were able to measure Na
abundances, and the circles mark the stars for which we were not able to
do so. The location of the non-member that was observed is marked 
with a square. The dashed circles represent the core, 
half-mass, and King truncation radius, respectively,
going from smallest to largest. The references for these radii are given in 
Table~1\label{fig3}.}
\end{figure}

\begin{figure}
\includegraphics*[width=0.5 \textwidth]{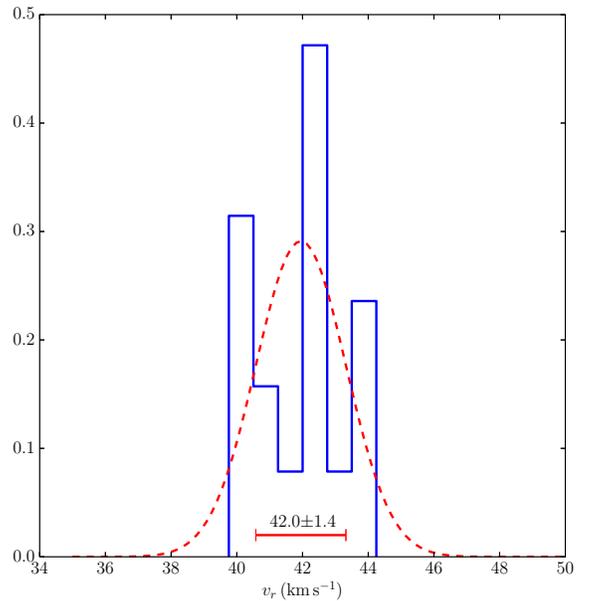}
\caption{The radial velocity (RV) distribution of the stars we observed with Hydra. The dashed line in the
plot is a normal distribution fit to the histogram shown. The single error bar is
centered at the cluster average RV, and the length of the error bar is the standard deviation of the RV sample. The average RV
value and standard deviation for the sample are given on top of the the error bar. 
In measuring the RVs in our sample, we found one field star that met
the photometric criteria, but was easy ruled out as a cluster member with an RV of
-27.1 $\mathrm{km \, s^{-1}}$\label{fig4}.}
\end{figure}
The stars for this study were selected based on their location on the $B$ vs $(B-V)$
CMD from \cite{1995AJ....109..269S}, as shown in Figure 2. Our initial sample only
included those stars that lie along the RGB in the CMD with a $V \leq 15.5$. 
At the spectral resolution and high S/N ratio needed to accurately measure 
abundances, stars fainter than this would require unrealistically long exposure times. 
The stars that met these criteria, and were observed with Hydra, are represented 
by open diamonds in Figure 2. A summary of the observational data for our sample
is given in Table 2. There we list the positions, photometric values, 
radial velocities, S/N of the spectra, and cluster membership of each star. 
The positions of the stars observed in this study are marked 
on a DSS image of NGC~5053 shown in Figure 3.

Radial velocities were determined for the observed sample using 
the \textit{fxcor} task in the NOAO package in IRAF. For each star we measured its radial velocity 
with four radial velocity standards observed at the beginning of each night. The 
typical standard deviation of the four individual measurements for a given star 
was 0.3 $\mathrm{km \, s^{-1}}$ and each measurement had an average error of 
2.2 $\mathrm{km \, s^{-1}}$, as estimated by \textit{fxcor}. We adopted the 
weighted mean of the four measurements as the final radial velocity value for each 
star, which is listed in Table 2. The error associated with each velocity in 
Table 2 is the error on the weighted mean. We also denote which stars are 
cluster members and which are non-members with a Y and N flag, respectively. 
The average radial velocity of cluster members is 
 42.0 $\pm 1.4 \, \mathrm{km \, s^{-1}}$. Our sample average radial velocity  
is in agreement with the previously determined value for the cluster 
$v_{r} = 42.6 \pm 0.3 \, \mathrm{km \, s^{-1}}$  \citep{2014arXiv1411.1763K}. 
The radial velocity distribution of the stars in our sample is shown in 
Figure 4. The only non-member we observed (Star 22), as listed in Table 2, 
is not included in Figure 4 because it sits too far away from the 
cluster distribution, and would skew the x-axis of the plot.

\begin{deluxetable*}{cccccccccccc}[!h]
\tabletypesize{\scriptsize}
\tablecaption{Observational Data \label{tbl-2}}
\tablewidth{\textwidth}
\tablehead{\colhead{ID}& \colhead{R.A.} & \colhead{Decl.} & \colhead{$V$} & \colhead{$M_{v}$} & \colhead{$(B-V)_{0}$} & \colhead{$(V-K)_{0}$} &  \colhead{$v_{r}$} & \colhead{$\sigma_{v_{r}}$} 
& \colhead{S/N} & \colhead{Member} \\ 
\colhead{ } & \colhead{J2000} & \colhead{J2000} &\colhead{ } & \colhead{ } & \colhead{ } & \colhead{ } &  \colhead{($\mathrm{km \, s^{-1}}$)} & \colhead{($\mathrm{km \, s^{-1}}$)} 
& \colhead{ } & \colhead{Y/N}}

\startdata
  2 & 13:16:26.71 & +17:41:01.6 &  14.03  & -2.23  & 1.09  & 2.71 & 44.2 & 0.8 &  83 & Y \\
  3 & 13:16:12.23 & +17:46:22.9 &  14.08  & -2.18  & 1.08  & 2.73 & 44.0 & 1.0 &  70 & Y \\
  4 & 13:16:20.60 & +17:42:46.4 &  14.45  & -1.81  & 0.99  & 2.58 & 42.2 & 0.8 &  82 & Y \\
  6 & 13:16:35.95 & +17:41:12.8 &  14.56  & -1.70  & 0.97  & 2.55 & 42.2 & 0.6 & 108 & Y \\
  9 & 13:16:26.45 & +17:40:07.6 &  14.68  & -1.58  & 0.89  & 2.41 & 43.8 & 1.1 & 108 & Y \\
 10 & 13:16:24.50 & +17:39:50.0 &  14.82  & -1.44  & 0.90  & 2.44 & 40.6 & 1.8 &  50 & Y \\
 11 & 13:16:57.64 & +17:38:05.1 &  14.83  & -1.43  & 0.89  & 2.43 & 40.5 & 0.7 &  71 & Y \\
 12 & 13:16:38.19 & +17:40:51.8 &  14.91  & -1.35  & 0.95  & 2.51 & 42.3 & 1.0 & 109 & Y \\
 14 & 13:16:37.90 & +17:45:35.5 &  15.04  & -1.22  & 0.88  & 2.39 & 40.1 & 1.2 &  80 & Y \\
 15 & 13:16:20.74 & +17:41:05.9 &  15.10  & -1.16  & 0.85  & 2.36 & 43.1 & 0.9 &  83 & Y \\
 16 & 13:16:25.56 & +17:44:02.3 &  15.15  & -1.11  & 0.77  & 2.20 & 42.6 & 1.4 &  79 & Y \\
 17 & 13:16:18.18 & +17:39:05.7 &  15.19  & -1.07  & 0.78  & 2.20 & 42.6 & 1.2 & 121 & Y \\
 18 & 13:16:32.51 & +17:42:17.4 &  15.22  & -1.04  & 0.77  & 2.16 & 41.7 & 1.1 &  60 & Y \\
 19 & 13:16:34.20 & +17:43:53.1 &  15.24  & -1.02  & 0.83  & 2.27 & 40.8 & 0.8 &  70 & Y \\
 20 & 13:16:22.27 & +17:41:53.6 &  15.25  & -1.01  & 0.83  & 2.53 & 42.1 & 0.8 &  82 & Y \\
 21 & 13:16:22.13 & +17:38:50.1 &  15.26  & -1.00  & 0.83  & 2.31 & 40.3 & 1.2 &  73 & Y \\
 22 & 13:16:28.55 & +17:41:12.6 &  15.35  & -0.91  & 0.81  & 1.82 &-27.1 & 0.7 &  80 & N \\
 23 & 13:16:33.91 & +17:42:22.9 &  15.37  & -0.89  & 0.80  & 2.26 & 39.7 & 1.4 &  64 & Y 
\enddata                                       

\end{deluxetable*}
\begin{deluxetable}{ccccc}
\tabletypesize{\scriptsize}
\tablecaption{Derived Atmospheric Parameters \label{tbl-3}}
\tablewidth{0.5\textwidth}
\tablehead{\colhead{ID}& \colhead{$T_{eff}$} & \colhead{$BC_{v}$} & \colhead{log(g)} & \colhead{$v_{t}$} \\ 
\colhead{ }& \colhead{K} & \colhead{ } & \colhead{$\mathrm{cm \, s^{-2}}$} & \colhead{$\mathrm{km \, s^{-1}}$}}
\startdata
  2  &  4460  & -0.53   & 0.80    & 1.67 \\
  3  &  4400  & -0.55   & 0.80    & 1.73 \\
  4  &  4500  & -0.48   & 1.10    & 1.67 \\
  6  &  4580  & -0.46   & 1.10    & 1.85 \\
  9  &  4660  & -0.42   & 1.20    & 1.85 \\
 10  &\nodata &\nodata  &\nodata  &\nodata \\
 11  &  4570  & -0.46   & 1.20    & 2.00 \\
 12  &  4490  & -0.51   & 1.20    & 2.08 \\
 14  &  4630  & -0.44   & 1.40    & 1.65 \\
 15  &  4630  & -0.43   & 1.40    & 2.09 \\
 16  &  4760  & -0.38   & 1.50    & 1.80 \\
 17  &  4750  & -0.39   & 1.50    & 1.41 \\
 18  &  4850  & -0.35   & 1.69    & 1.34 \\
 19  &  4650  & -0.43   & 1.40    & 1.65 \\
 20  &  4700  & -0.33   & 1.50    & 2.11 \\
 21  &  4680  & -0.41   & 1.50    & 1.19 \\
 22  &\nodata &\nodata  &\nodata  &\nodata\\
 23  &  4700  & -0.41   & 1.50    & 1.50 \\
\enddata                                       
\end{deluxetable}

\section{Analysis}

\subsection{Atmospheric Parameters}
Initial estimates of the atmospheric parameters ($T_{eff}, BC_{v}$) were 
determined from the $B,V,I$ photometry \citep{1995AJ....109..269S} and 2MASS 
$J,H,K$ for each star using the relations found in \cite{1999A&AS..140..261A}, 
using the $(B-V)$, $(V-K)$, $(J-H)$, and $(J-K)$ colors. The colors were 
dereddened with an E(B-V) = 0.04 \citep{2006ApJ...651L..33L}, which is a 
consistent value across the literature, following the relations found by 
\cite{1985ApJ...288..618R}. Surface gravities were calculated based on 
the derived effective temperatures and bolometric corrections assuming a mass of 
$0.8M_{\odot}$ for every star in the sample and a distance modulus of 
$(m-M)_{V} = 16.26$ \citep{2007AJ....133.1658S}. Due to the limited number of
measurable Fe I and Fe II lines in our spectral region, it was not possible to 
adjust the $T_{\mathrm{eff}}$ and log($g$) from their photometrically determined values
using excitation or ionization equilibrium.

The microturbulent velocity ($v_{t}$) of each star was determined using the 
relationships found in \cite{2004A&A...422..951C}, \cite{2008ApJ...681.1505J}, 
and \cite{1996AJ....111.1689P}. The initial estimate of $v_{t}$ was an average 
of the velocities resulting from the relations found in the three references 
listed above. The $v_{t}$ was later adjusted to minimize the dependence of 
derived [Fe/H] abundance on Fe I line strength. The final adopted atmospheric 
parameters are listed in Table 3. MARCS LTE model atmospheres 
\citep{2008A&A...486..951G} were interpolated to the photometrically determined 
atmospheric parameters in order to be used in the abundance measurements. 
We did not derive these stellar parameters for Star 10 because its spectrum was 
too noisy to use for abundances measurements. Star 22 is also listed without 
these parameters because it was determined to not be a cluster member 
from its radial velocity.

\subsection{Equivalent Width (EW) Measurements}

The iron (Fe), nickel (Ni), titanium (Ti), and calcium (Ca) abundances in our 
sample were determined from EW measurements of individual Fe I, 
Ni I, Ti I and Ca I lines, respectively. The spectra in our sample were compared 
to a high resolution solar spectrum \citep{2011ApJS..195....6W} 
to confirm the location of lines and possible blends. EWs were measured using 
the \textit{splot} task in IRAF on the continuum-normalized spectra. 
The final line lists were complied with log($gf$) values and excitation potentials 
taken from the \textit{Gaia}-ESO Survey (Heiter et al. in prep). The EW measurements and 
atomic parameters for all lines used are available electronically. 
The abundances were determined using the \textit{abfind} task in the 2010 August 
version of MOOG \citep{1973ApJ...184..839S}.
We assumed the solar abundances as found by \cite{1989GeCoA..53..197A}.

\begin{deluxetable}{ccccc}


\tablecaption{Equivalent Widths}
\tablehead{\colhead{Element} & \colhead{$\lambda$} & \colhead{log($gf$)} & \colhead{EP} & \colhead{EW} \\ 
\colhead{} & \colhead{(\AA)} & \colhead{} & \colhead{(eV)} & \colhead{(m\AA)} } 
\tablewidth{0.5\textwidth}
\startdata
Fe I & 6136.61 & -1.402 & 2.45 & 99.9 \\
Fe I & 6136.99 & -2.950 & 2.20 & 32.2 \\
Fe I & 6137.69 & -1.402 & 2.59 & 999.9 \\
Fe I & 6151.62 & -3.295 & 2.18 & 21.6 \\
Fe I & 6173.33 & -2.880 & 2.22 & 36.1 
\enddata

\tablecomments{Table 4 is published in its entirety in the electronic edition of the {\it Astrophysical Journal}.  
A portion is shown here for guidance regarding its form and content. The full version contains all the lines for the other 17 stars.}

\end{deluxetable}

\subsection{Spectral Synthesis}

\begin{figure*}
\centering
\begin{minipage}[b]{\linewidth}
\includegraphics*[width=0.5 \textwidth]{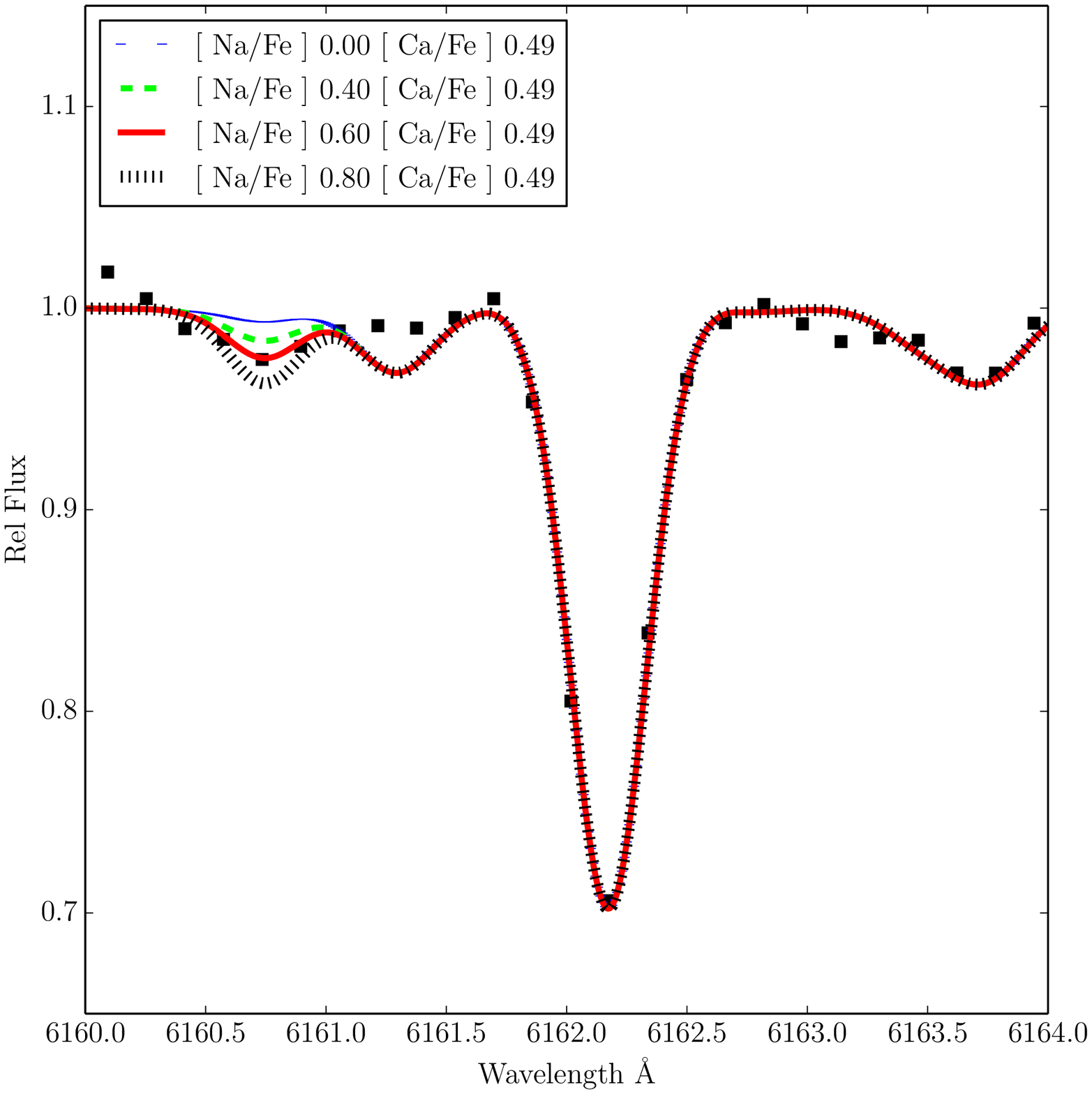}
 \includegraphics*[width=0.5 \textwidth]{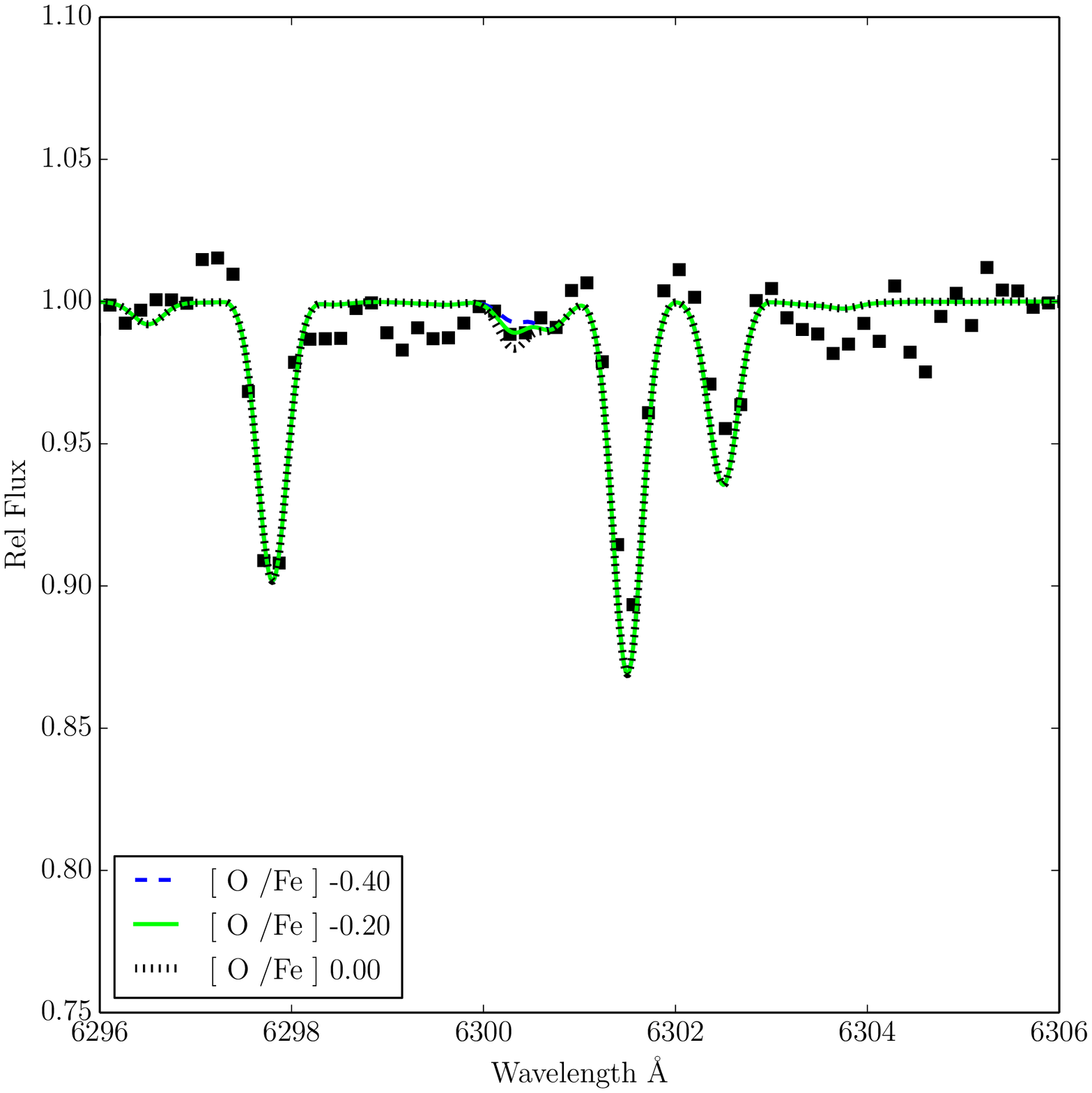}
\end{minipage}

\caption{An example of the spectral synthesis 
performed to determine the Na and O abundances 
for Star 6. Left panel: the spectral region 
around the Na I line near 6160 \AA. The strong 
line located at approximately 6162 \AA \, is a Ca I line. 
Right panel: the spectral window near the [O I] line near
6300 \AA. In each panel the data are plotted as black squares. 
The spectrum in the right hand panel has been corrected for telluric features.\label{fig5} }
\end{figure*}

\begin{figure*}
\begin{minipage}[b]{\linewidth}
\includegraphics*[width=0.5 \textwidth]{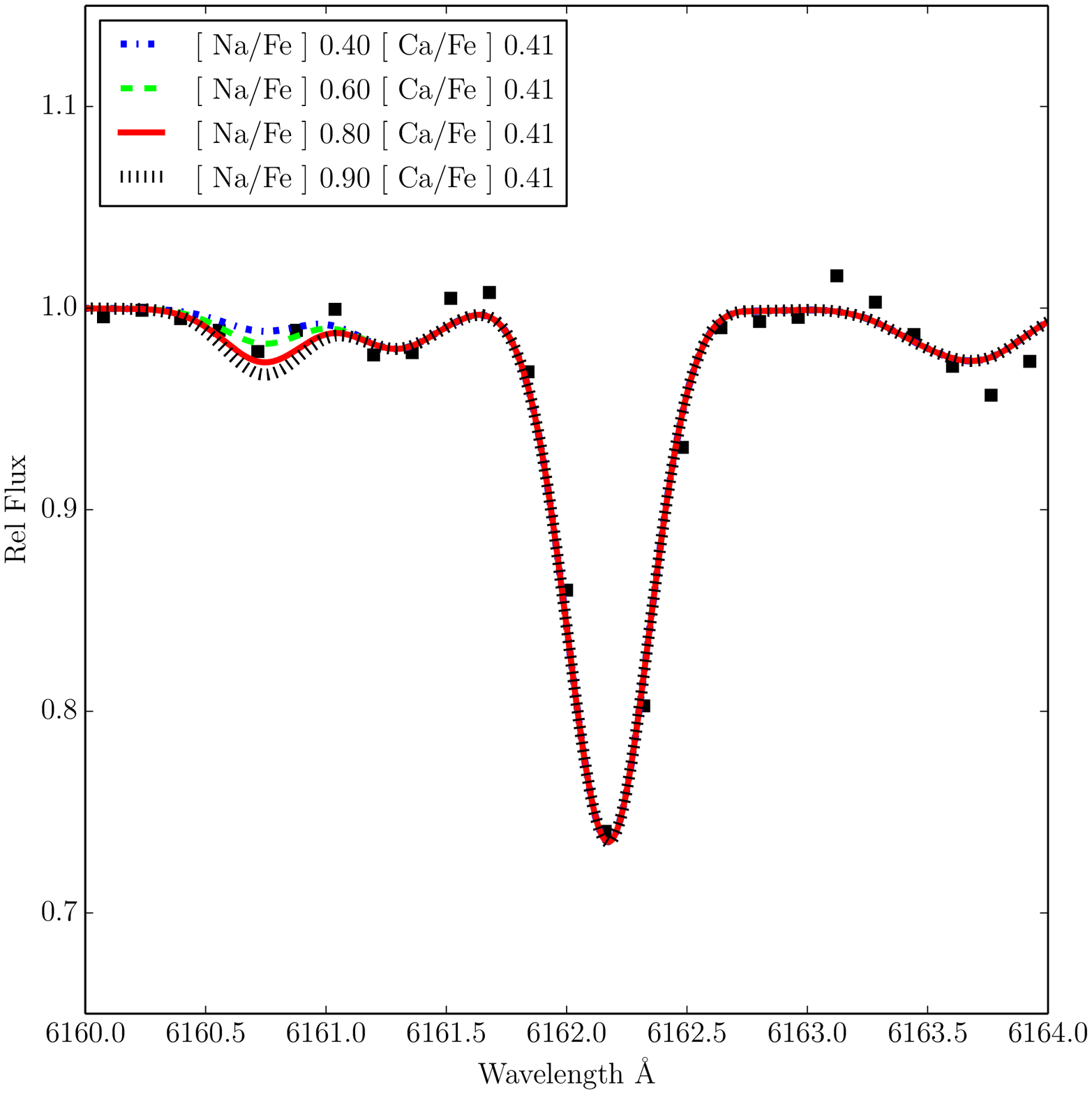}
 \includegraphics*[width=0.5 \textwidth]{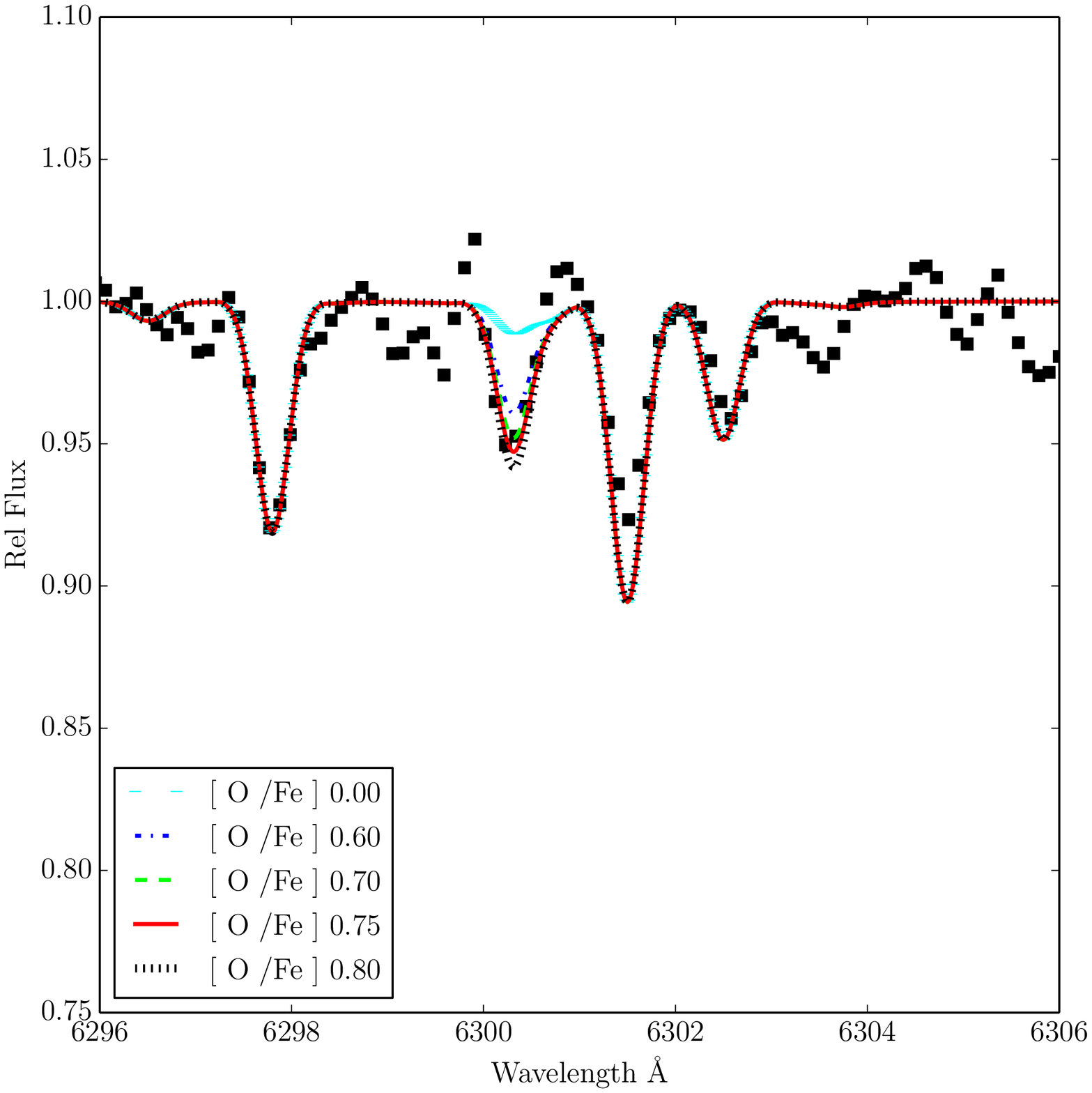}
\end{minipage}
\caption{As in Figure 5, but for Star 12 with similar atmospheric parameters
but very different oxygen abundance.\label{fig6} }
\end{figure*}

The  sodium (Na), oxygen (O), and barium (Ba) abundances were determined using 
spectral synthesis to account for blends, molecular bands,
and in the case of Ba, hyperfine structure and the isotopic mix of the element. Synthetic spectra were 
generated using the \textit{synth} task in MOOG, with MARCS model
atmospheres that matched the atmospheric parameters of each star, and an 
[Fe/H] abundance as determined through EW measurements.
Assuming Gaussian broadening, the FWHM of the synthetic spectral lines was set 
to match the FWHM of unblended lines in the observed spectra. 
The final abundances were determined by creating synthetic spectra with a range 
of abundances for the species and line in question, 
and choosing the synthesis that best matched the observed spectra. 
An error was assigned to these abundances 
by determining how much the abundance could be varied before the 
synthetic spectrum would greatly deviate from the observed lines.  Examples
of the spectral synthesis are shown in Figure 5 and Figure 6.

In our spectral region, there are two Na I lines that we used
to determine the [Na/Fe] abundance at $6154$ \AA \, and $6160$ \AA. For many of the stars,
the line at $6154$ \AA \, was too weak to measure and the final 
[Na/Fe] abundances were based on the single line at $6160$ \AA. In the cases where
both lines were present, an average of the [Na/Fe] resulting from 
each line was used as the final abundance. There were also cases where both of
these lines were too noisy to determine an [Na/Fe] abundance for 
the star. In total, we were able to determine the [Na/Fe] abundance in 11 of the 18
stars observed.

Oxygen abundances were determined from the 
[O I] line at $6300$ \AA. In this region of the spectrum there is a significant amount of contamination 
from telluric features. To remove these features, we divided 
the observed spectrum by the scaled spectrum of a hot B star, 
which acted as our template for the telluric features. The synthetic 
spectra were then compared to these telluric corrected spectra when
determining the final [O/Fe] abundances. As a test, we also determined 
the [O/Fe] abundances from uncorrected spectra and there were no differences
in the results. This indicated to us that the [O I] line was not significantly 
affected by the neighboring telluric lines. There is also a strong night 
sky emission line near this [O I] feature. 
The radial velocity shift of the NGC~5053 stars moved the sky line off of the [O I] feature. 
In some cases, however, the strength of the sky line varied enough so that the
sky subtraction was not good enough to completely remove the line. In these 
cases the residual from the sky line was large enough to affect the [O I]
feature. It is for that reason that we were only able to 
measure [O/Fe] abundances in 8 stars.

The [Ba/Fe] abundances were determined from the Ba II line at $6141$ \AA. In order 
to create an accurate synthetic spectrum including this feature,
we had to assume an isotopic mix of Ba. This is complicated by the fact that Ba is both 
an \textit{s-} and \textit{r-}process neutron capture element. 
At low metallicities, such as those in NGC~5053, the Ba production is thought to 
be dominated by the \textit{r-} process because of the slower timescales over which
the \textit{s-}process would contribute to the overall Ba production. Following 
the work of \cite{1998AJ....115.1640M}, we only considered the following 
isotopes $^{135}\mathrm{Ba},^{137}\mathrm{Ba},^{138}\mathrm{Ba}$, with a ratio 
of $40:32:28$, respectively. We also adopt the \textit{r-}process 
solar composition of $\mathrm{log}(\epsilon_{\odot}(\mathrm{Ba})_{r}) = 1.46$, 
as found by \cite{1998AJ....115.1640M}. It should be noted that there
is an Fe I line at $6140$ \AA \, which can blend with the Ba II line, resulting in 
inaccurate [Ba/Fe] abundance determinations. At the metallicity of NGC~5053,
however, the line is very weak so the effect is minimal.

\subsection{Uncertainties}

To quantify the errors in our abundance measurements, we determined
how much the abundances changed with variations in the atmospheric
parameters of a given model. New models were generated with one of 
the atmospheric parameters varying, while the others were held constant.
The abundance analysis was then repeated with these new models 
until we had sampled a sufficient grid of the parameter space to assess the
errors induced by their individual uncertainties. A summary 
of this error analysis is given in Table 5. The analysis was performed 
on a star with the following atmospheric parameters over the range indicated 
with each value: $T_{eff} = 4760 \pm 100$ (K),  
$\mathrm{log}$ $g$, = 1.50 $\pm 0.2$ (cgs), [M/H] = -2.5 $\pm 0.2$ (dex),$v_{t}$ = 1.8 $\pm 0.2 \,
\mathrm{km \, s^{-1}}$. 

In Table 5 we list the $\sigma_{\mathrm{obs.}}$ and $\sigma_{\mathrm{total}}$ for each abundance measured.
For abundances determined using EWs, $\sigma_{\mathrm{obs.}}$ 
is the standard deviation of abundances determined from 
the number of lines listed in the table. 
In cases where only one line was measured this field 
is left blank. For those abundances that were determined with
spectral synthesis (Na, O, Ba), $\sigma_{\mathrm{obs.}}$ 
is the error assigned to each abundance as described in the previous section.
To calculate $\sigma_{\mathrm{total}}$, we added the 
uncertainties from the atmospheric parameters in quadrature. 
From this analysis we see that our observed uncertainty is 
consistent with that expected from variations in atmospheric parameters.

\begin{deluxetable*}{lccccccc}[t!]
\tabletypesize{\scriptsize}
\tablecaption{Uncertainties on Atmospheric Parameters\label{tbl-4}}
\tablewidth{\textwidth}
\tablehead{\colhead{Ion}& \colhead{$T_{eff} \, \pm 100$} & \colhead{log $g  \, \pm 0.20$} & \colhead{[M/H] $\, \pm 0.10$} & \colhead{$v_{t}\, \pm 0.25$} 
& \colhead{$\sigma_{\mathrm{total}}$} & \colhead{No.} & \colhead{$\sigma_{\mathrm{obs}}$}\\
\colhead{ }& \colhead{(K)} & \colhead{(cgs)} & \colhead{(dex)} & \colhead{($\mathrm{km \, s^{-1}}$)} & \colhead{(dex)} & \colhead{Lines}
& \colhead{(dex)}}
\startdata

Fe I              &  $\pm 0.13$  &  $\pm 0.02$  &  $\pm 0.02$  &  $\mp 0.03$  & 0.14 & 9  &  0.07       \\
$[\mathrm{O  I}]$ &  $\pm 0.10$  &  $\pm 0.07$  &  $\pm 0.02$  &  $\pm 0.00$  & 0.12 & 1  &  0.20       \\
Na I              &  $\pm 0.10$  &  $\pm 0.07$  &  $\pm 0.02$  &  $\mp 0.05$  & 0.13 & 1  &  0.20       \\
Ca I              &  $\pm 0.11$  &  $\pm 0.03$  &  $\pm 0.01$  &  $\mp 0.07$  & 0.13 & 3  &  0.09       \\
Ti I              &  $\pm 0.12$  &  $\pm 0.01$  &  $\pm 0.01$  &  $\mp 0.01$  & 0.12 & 1  & \nodata     \\
Ni I              &  $\pm 0.13$  &  $\pm 0.01$  &  $\pm 0.01$  &  $\mp 0.01$  & 0.13 & 1  & \nodata     \\
Ba II             &  $\pm 0.07$  &  $\pm 0.05$  &  $\pm 0.05$  &  $\mp 0.05$  & 0.11 & 1  &  0.20       

\enddata
\end{deluxetable*}

\section{Results}
\subsection{Fe, Ca, Ti, Ni, Ba}

\begin{figure*}
\includegraphics[width=\textwidth]{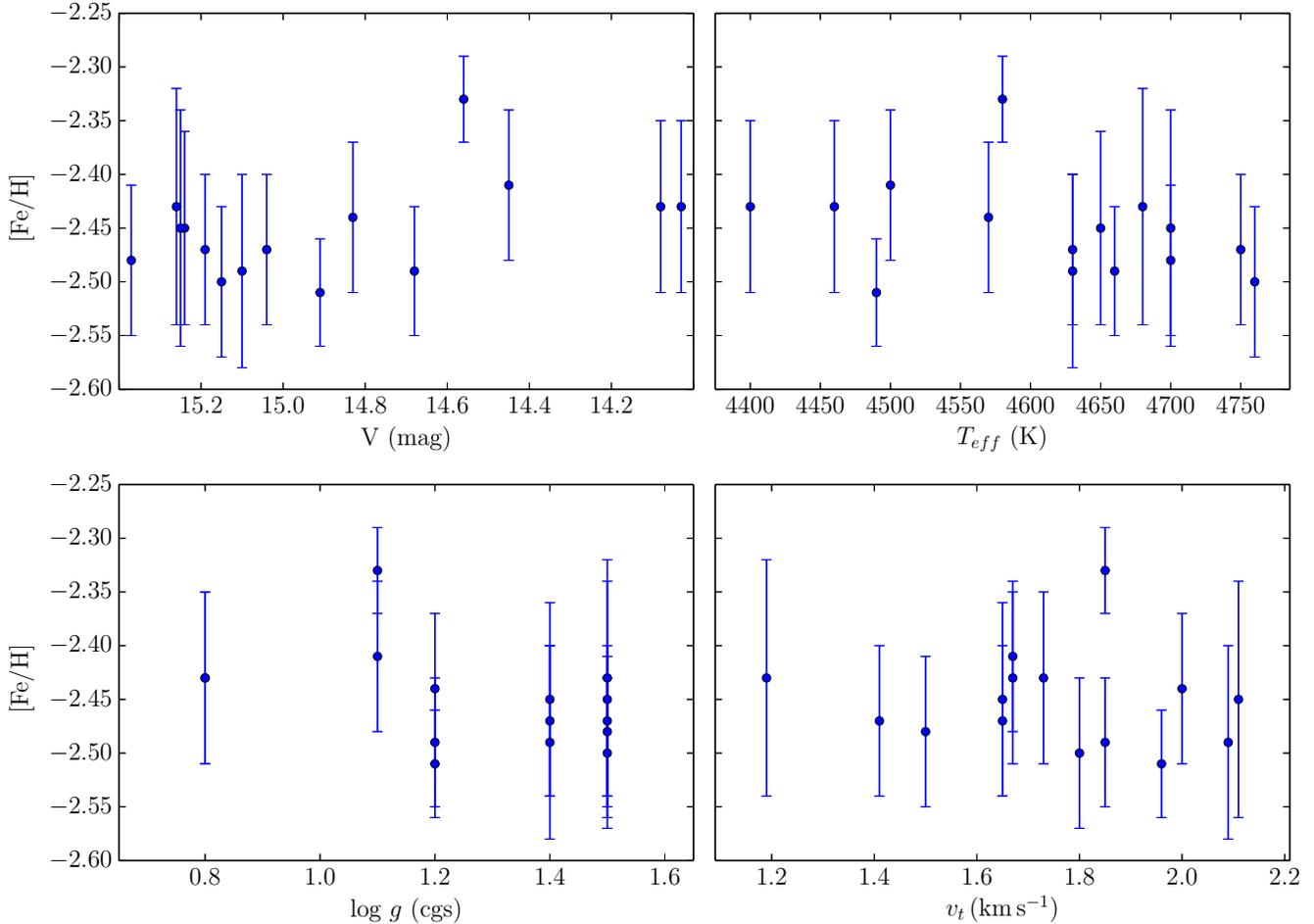}
\caption{[Fe/H] as a function of apparent magnitude $V$, $T_{\mathrm{eff}}$, log($g$), and $v_{t}$. The vertical 
error bars in each of the plots represent the standard deviation of the 
[Fe/H] abundance of a given star. \label{fig7} }
\end{figure*}

In Figure 7 we plot the [Fe/H] abundance as a function of V, $T_{\mathrm{eff}}$, log($g$), and $v_{t}$
for each star to see if there are trends in the measured [Fe/H] values with these parameters.
The error bars in the vertical direction represent the standard deviation of the [Fe/H] values given by
the Fe I lines in a given spectrum. We do not find any trends in [Fe/H] 
with any of these parameters.

The individual stellar abundances for Fe, Ca, Ti, Ni, and Ba are given in Table 6 
with the number of lines averaged together to reach the final abundance. 
In the second to last row of Table 6
we list the the average cluster abundance for Fe, Ca, Ti, Ni, and Ba 
with the number of stars used in the average and the standard deviation of the measurements.
The final row of Table 6 lists the assumed solar abundances used to calculate
the abundance ratios given in this table.
In Figure 8 we present a box- and -whisker plot of the Ca, Ti, Ni and Ba abundances 
to illustrate the median cluster abundance of these species, 
as well as their individual distributions in the cluster.

\begin{deluxetable*}{cccccccccccccccc}

\tabletypesize{\scriptsize}
\tablecaption{Fe, Ca, Ti, Ni, and Ba Abundances in NGC~5053\label{tbl-6}}
\tablehead{\colhead{ID} & \colhead{[Fe/H]} & \colhead{$\sigma_{\mathrm{Fe}}$} & \colhead{n} & \colhead{[Ca/Fe]} & \colhead{$\sigma_{\mathrm{Ca}}$} & \colhead{n} 
& \colhead{[Ti/Fe]} & \colhead{$\sigma_{\mathrm{Ti}}$} & \colhead{n} & \colhead{[Ni/Fe]} & \colhead{$\sigma_{\mathrm{Ni}}$} & \colhead{n} & \colhead{[Ba/Fe]}
 & \colhead{$\sigma_{\mathrm{Ba}}$} & \colhead{n}}
\startdata
  2  & -2.43  &   0.08  &  17    & 0.60    &   0.06   &   5    &    0.49    & \nodata &    1   &  -0.06  & 0.15   &    2   &   -0.58  &    0.1   &     1   \\
  3  & -2.43  &   0.08  &  11    & 0.46    &   0.01   &   2    &    0.32    &  0.11   &    3   & \nodata & \nodata&    1   &   -0.33  &    0.1   &     1   \\
  4  & -2.41  &   0.07  &  11    & 0.43    &   0.00   &   2    &    0.09    &  0.14   &    2   &   0.07  & 0.31   &    2   &   -0.58  &    0.2   &     1   \\
  6  & -2.33  &   0.04  &  10    & 0.49    &   0.04   &   4    &    0.26    &  0.02   &    2   &  -0.26  & 0.02   &    2   &   -0.58  &    0.2   &     1   \\
  9  & -2.49  &   0.06  &  12    & 0.44    &   0.07   &   6    &  \nodata   & \nodata & \nodata&  -0.07  &\nodata &    1   &   -0.58  &    0.2   &     1   \\
 10  &\nodata & \nodata &\nodata & \nodata &  \nodata & \nodata&  \nodata   & \nodata & \nodata& \nodata &\nodata &\nodata &  \nodata &  \nodata & \nodata \\
 11  & -2.44  &   0.07  &   8    & 0.32    &   0.07   &   3    &    0.28    &  0.22   &    2   &   0.02  & \nodata&    1   &   -0.58  &    0.1   &     1   \\
 12  & -2.51  &   0.08  &  13    & 0.41    &   0.07   &   4    &    0.19    &  0.07   &    2   &   0.02  & \nodata&    1   &   -0.53  &    0.2   &     1   \\
 14  & -2.47  &   0.07  &   9    & 0.51    &   0.14   &   5    &    0.39    & \nodata &    1   &   0.15  & \nodata&    1   &   -0.78  &    0.2   &     1   \\
 15  & -2.48  &   0.09  &  13    & 0.32    &   0.05   &   2    &    0.32    &  0.20   &    2   &  -0.15  & \nodata&    1   &   -0.58  &    0.2   &     1   \\
 16  & -2.50  &   0.07  &   9    & 0.51    &   0.09   &   3    &    0.56    & \nodata &    1   &   0.3   &\nodata&     1   &   -0.43  &    0.2   &     1   \\
 17  & -2.47  &   0.07  &   9    & 0.42    &   0.09   &   6    &  \nodata   & \nodata & \nodata& \nodata &\nodata &\nodata &   -0.58  &    0.1   &     1   \\
 18  &\nodata & \nodata &\nodata & \nodata &\nodata & \nodata &   \nodata   & \nodata & \nodata& \nodata &\nodata &\nodata &  \nodata &  \nodata & \nodata \\
 19  & -2.45  &   0.09  &  11    & 0.62    &   0.06   &   2    &    0.32    &  0.15   &    2   &   0.28  &\nodata &    1   &   -0.43  &    0.1   &     1   \\
 20  & -2.45  &   0.11  &  13    & 0.43    &   0.12   &   3    &  \nodata   & \nodata & \nodata&   0.15  &\nodata &    1   &   -0.53  &    0.1   &     1   \\
 21  & -2.43  &   0.11  &  13    & 0.5     &   0.16   &   5    &    0.27    & \nodata &    1   &   0.13  &\nodata &    1   &   -0.53  &    0.1   &     1   \\
 22  &\nodata & \nodata &\nodata & \nodata &\nodata & \nodata  &  \nodata   & \nodata & \nodata& \nodata &\nodata &\nodata &  \nodata &  \nodata & \nodata \\
 23  & -2.48  &   0.07  &   8    & 0.39    &   0.18   &   4    &    0.43    &  0.21   &    2   &   0.00  & 0.01   &    2   &   -0.43  &    0.1   &     1   \\
Cluster Avg.& -2.45  &   0.04  &   15    & 0.45    &   0.08   &   15    &    0.32    &  0.11   &    12   &   0.03 &   0.14   &    13  &   -0.54 & 0.1 & 15  \\
$\mathrm{log}(\epsilon_{\odot})$ & 7.52& & & 6.36    & &   &    4.99    &   &     &   6.25 &     &    &   2.13 &  & 
\enddata
\end{deluxetable*}

\begin{figure}
\includegraphics[width = 0.5 \textwidth]{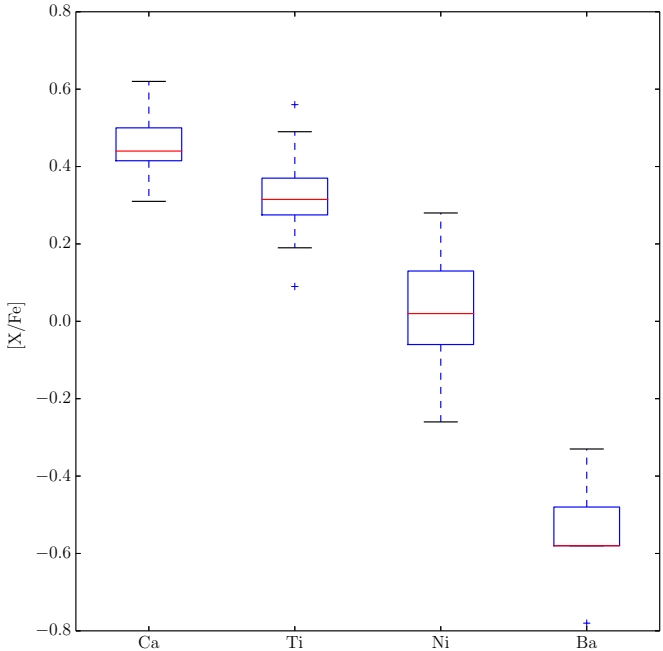}
\caption{Box- and- whisker plot of the Ca, Ti, Ni and Ba abundances in NGC~5053. 
The red line is the median of each abundance, the bottom of each box is the first quartile, the top of
each box is the third quartile, and the caps represent the minimum and maximum values.\label{fig8} }
\end{figure}

\begin{deluxetable}{ccccc}
\tabletypesize{\scriptsize}
\tablecaption{Na and O Abundances in NGC~5053\label{tbl-7}}
\tablewidth{0.45\textwidth}
\tablehead{\colhead{ID} & \colhead{$[\mathrm{Na/Fe}]$} & \colhead{Error} & \colhead{$[\mathrm{O/Fe}]$} & \colhead{Error} }
\startdata
   2  &  0.7      &  0.2      &  0.2      &  0.2      \\
   3  &  $<0.0$   &  \nodata  &  0.3      &  0.2      \\
   4  &  \nodata  &  \nodata  &  0.6      &  0.2      \\
   6  &  0.6      &  0.20     & -0.2      &  0.2      \\
   9  &  $<0.0$   & \nodata   &  0.4      &  0.2      \\
  10  &  \nodata  &  \nodata  &  \nodata  &  \nodata  \\
  11  &  1.0      &  0.2      &  \nodata  &  \nodata  \\
  12  &  0.6      &  0.2      &  0.75     &  0.1      \\
  14  &  0.8      &  0.3      &  \nodata  &  \nodata  \\
  15  &  \nodata  &  \nodata  &  \nodata  &  \nodata  \\
  16  &  0.9      &  0.3      &  $<0.2$   &  \nodata  \\
  17  &  0.0      &  0.2      &  0.4      &  0.2      \\
  18  &  \nodata  &  \nodata  &  \nodata  &  \nodata  \\
  19  &  \nodata  &  \nodata  &  \nodata  &  \nodata  \\
  20  &  0.9      &  0.2      &  \nodata  &  \nodata  \\
  21  &  0.8      &  0.2      &  \nodata  &  \nodata  \\
  22  &  \nodata  &  \nodata  &  \nodata  &  \nodata  \\
  23  &  \nodata  &  \nodata  &  \nodata  &  \nodata  
\enddata
\end{deluxetable}

\begin{figure}
\includegraphics[width=0.5 \textwidth]{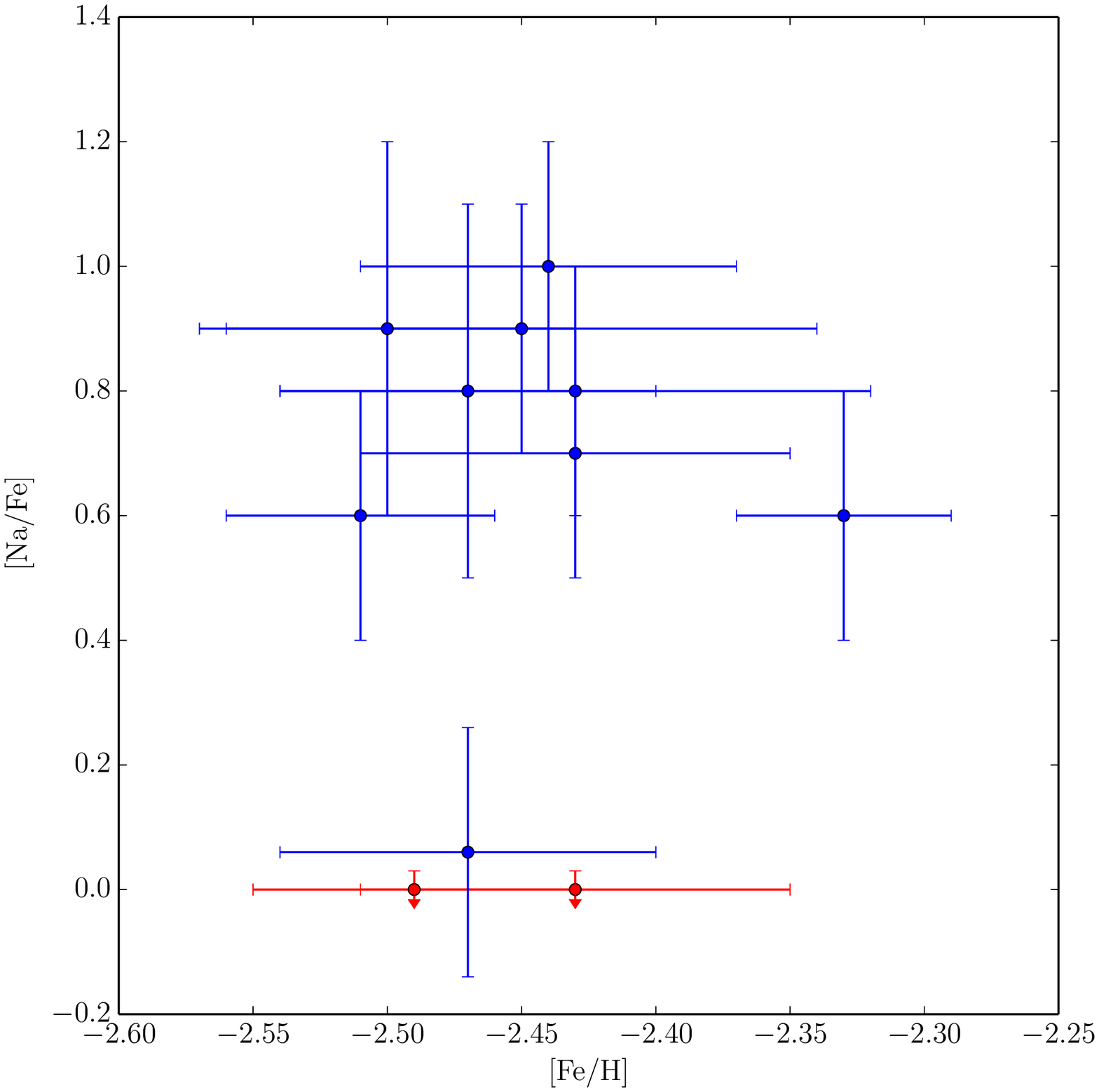}
\caption{The individual [Na/Fe] abundances for each star plotted vs their 
respective [Fe/H] abundance. This plot shows that there are two separate groups of [Na/Fe]
abundances present in our sample.\label{fig9} }
\end{figure}

\subsection{Na and O}

Using spectral synthesis we were able to determine 
the Na abundance in 11 of the 18 stars in our sample, and
the O abundance in 8 of them. There were 7 stars which 
had both Na and O abundances measured. These abundances
are summarized in Table 7.
The [Na/Fe] versus [Fe/H] abundance for each star and the [Na/Fe] distribution 
for the cluster are plotted in Figure 9 and Figure 10,
respectively. In Figure 10, the solid black line is a generalized 
histogram of the [Na/Fe] values. The generalized histogram was created
by treating each measurement as a Gaussian with a standard deviation 
equal to the error on that measurement.
These individual Gaussians were then added together to produce the 
distribution shown in Figure 10 by the solid line.
The distribution represented by this generalized histogram is best 
fit by two Gaussians centered at [Na/Fe] = --0.03 dex
and [Na/Fe] = 0.78 dex, with standard deviations of 0.30 and 0.28, respectively.
From these figures we clearly see that there is a spread on the order 
of 0.8 dex in the [Na/Fe] distribution of our sample, which is 
larger than the expected errors on the measurements.

\section{Discussion}


\begin{figure}
\epsscale{0.95}
\includegraphics[width=0.5\textwidth]{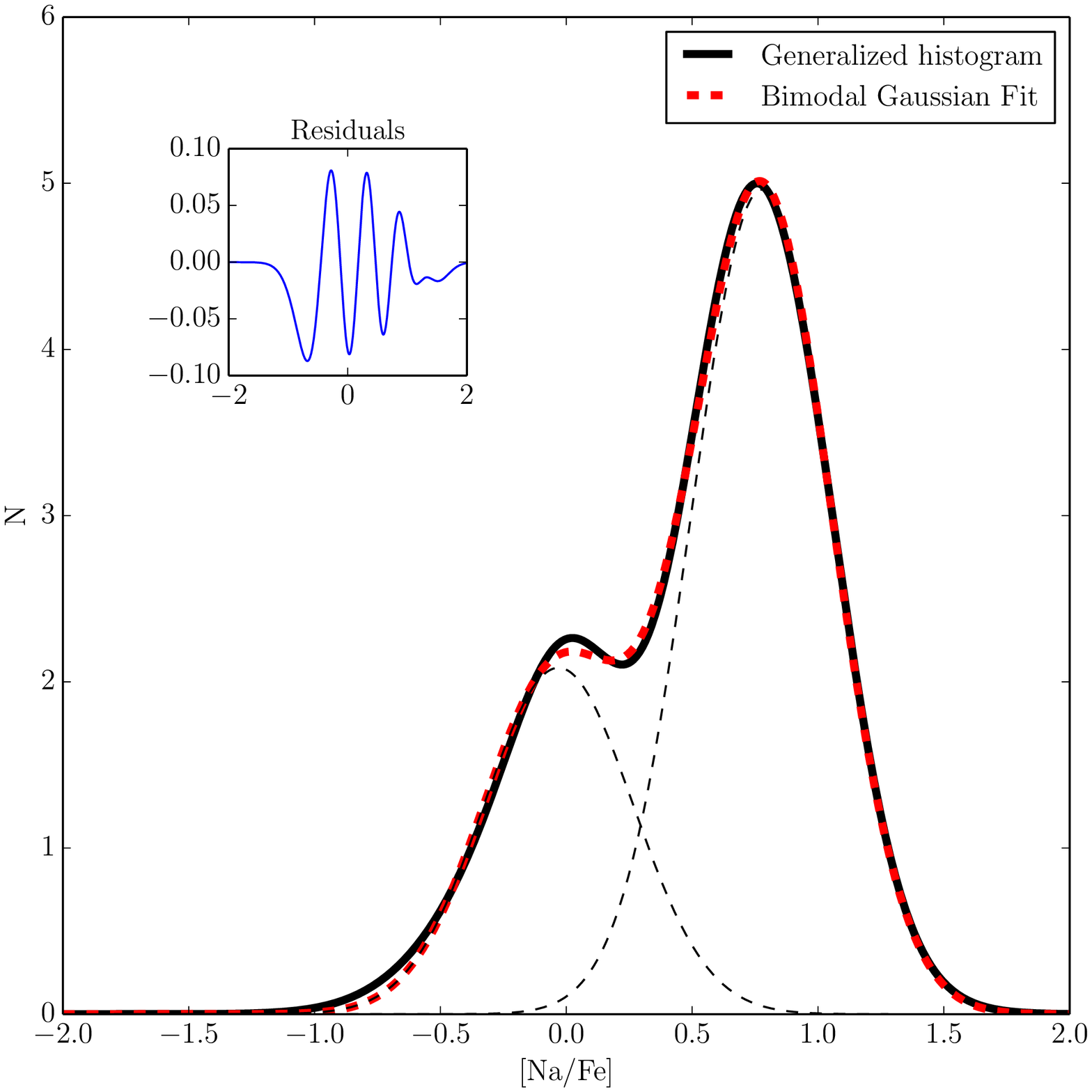}
\caption{Generalized histogram of the [Na/Fe] 
abundances. 
The thick dashed line is the two-peaked Gaussian fit to this distribution, with the 
individual Gaussian components plotted as thin dashed lines. 
The residuals between the fit and the generalized histogram are shown in the upper left-hand corner.\label{fig10} }
\end{figure}

\begin{figure*}
\epsscale{1.0}
\includegraphics[width=\textwidth]{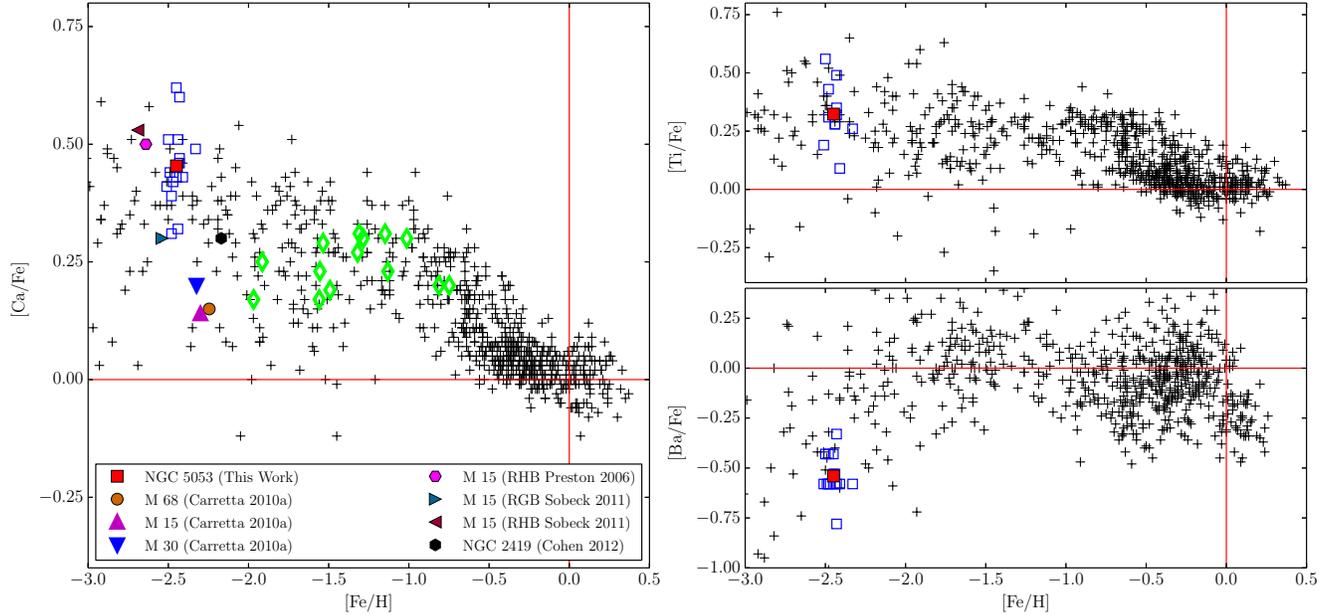}
\caption{The Ca, Ti, and Ba abundances of the stars in our sample plotted with a 
Milky Way Sample taken from \cite{2004AJ....128.1177V}, plotted as black crosses. 
The open squares in each plot represent the individual abundances for the individual
stars in our sample, the solid red square marks the average of these individual abundances.
Left panel: the open diamonds are Milky Way GC abundances 
as measured by \cite{2010ApJ...712L..21C} for Ca. Also on this plot, are additional measurements
for M15, with references given in the legend. While M15, M68, and M30 are from the
\cite{2010ApJ...712L..21C} sample, they are given different markers to highlight their values
in relationship to NGC~5053.}
\end{figure*}

\begin{figure}
\includegraphics[width=0.5\textwidth]{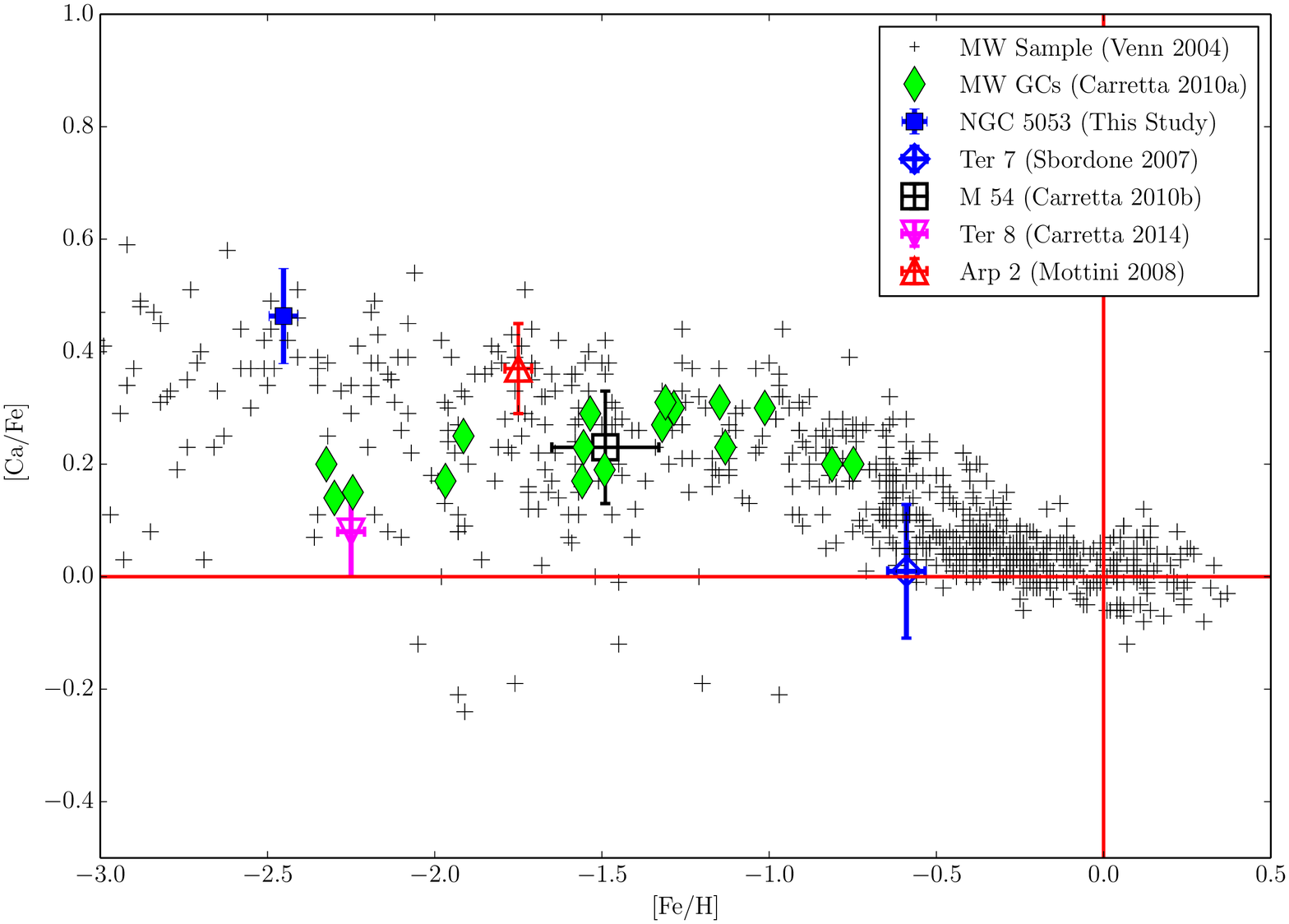}
\caption{A comparison of the NGC~5053 [Ca/Fe] abundance with the GCs traditionally thought to be associated with the Sgr dSph: 
M54, Arp 2, Terzan 7, and Terzan 8. \label{fig12} }
\end{figure}

\subsection{Comparison With MW}

With the abundances that we determined, we can 
place NGC~5053 in the context of the 
chemical composition of stars in the MW.
To make this comparison, in Figure 11 we plot the [Ca/Fe], 
[Ti/Fe], and [Ba/Fe] abundances of NGC~5053 with a MW sample taken from 
\cite{2004AJ....128.1177V}. In these plots the \cite{2004AJ....128.1177V} 
sample is plotted as black crosses. For each species,
the individual cluster members of NGC~5053 are plotted as 
open squares and the cluster average for that species is plotted as a filled square.
From these plots we can see that NGC~5053 lies within the envelope defined 
by MW stars at a similar metallicity. Specifically,
the alpha-elements in our sample, [Ca/Fe] 
and [Ti/Fe], are enhanced by typical amounts while the [Ba/Fe] 
is depleted. The large spread in the [Ti/Fe] abundances
comes from the fact that the TiI lines were relatively weak and in many cases
only 1 or 2 lines were available in our spectral window. 

On the [Ca/Fe] vs [Fe/H] plot we include
the [Ca/Fe] abundances for other GCs in the MW.
The open green diamonds mark the abundances as found
by \cite{2010ApJ...712L..21C}. The abundances for M68,
M15, and M30 from the \cite{2010ApJ...712L..21C} sample
are given different markers to highlight the locations of the
three most metal-poor GCs in their sample. In addition to the
their sample, we include
two other measurements of M15 from \cite{2006AJ....132...85P} and \cite{2011AJ....141..175S},
and the abundances for NGC~2419 as found by \cite{2012ApJ...760...86C}. The abundance ratios 
from other studies have been adjusted to our assumed solar values.

From the [Ca/Fe] panel in Figure 11, it would appear that NGC~5053 is more similar to
the MW field population than the GC sequence populated by the \cite{2010ApJ...712L..21C} sample. 
It is difficult, however, to know if this offset is real or caused by the systematic differences between
different studies. The abundances plotted for M15 illustrate this point well. Between the four measurements
for M15 on this panel, [Fe/H] ranges from -2.3 to -2.64 and [Ca/Fe] ranges from 0.11 to 0.53. In
the case of \cite{2011AJ....141..175S} the difference in abundances could arise from the evolutionary 
state of the stars being studied, but this could not account for the larger offsets shown in the plot.
Potential systematic differences in abundance analyses 
can arise from a multitude of effects, including different measurement and analysis techniques, 
adopted stellar models, stellar atmospheric parameters, and atomic parameters. The samples
collected to make this plot did have a few Ca I and Fe I lines in common with our work and these
had identical log($gf$) values, but most lines were unique to each study. With these considerations in mind,
we can say that NGC~5053 exhibits alpha enhancement that is typical of MW field stars and GCs at
a similar metallicity and is among the most metal-poor GCs in the MW.

\subsection{Comparison with Sgr dSph GCs}
Throughout the literature, M54, Arp 2, Terzan 7, and Terzan 8 
are considered to be GCs that originated from the Sgr dSph, with M54 being at the nucleus
of the dwarf galaxy \citep[see][]{1994Natur.370..194I,1995AJ....109.2533D,2010ApJ...718.1128L}.
In Figure~12 we plot the [Ca/Fe] of NGC~5053
along with these four clusters versus their respective [Fe/H] abundances. Also plotted is the MW
star sample from \cite{2004AJ....128.1177V}. In this figure all 
of the abundances have been adjusted to our assumed solar values.
As mentioned earlier and demonstrated in Figure 11, 
the apparent offset of some of the Sgr dSph GCs from the MW GC sequence 
could be due to the systematic differences between the studies. In general, the GCs
considered to be members of the Sgr dSph do not show [Ca/Fe] values that are drastically different from
either the MW GC sequence or the field stars at their respective metallicities, 
despite the systematic differences between studies.
If NGC~5053 were to be confirmed as a member of the 
Sgr dSph it would be the most metal-poor cluster known to date.

\begin{figure}
\epsscale{1.0}
\plotone{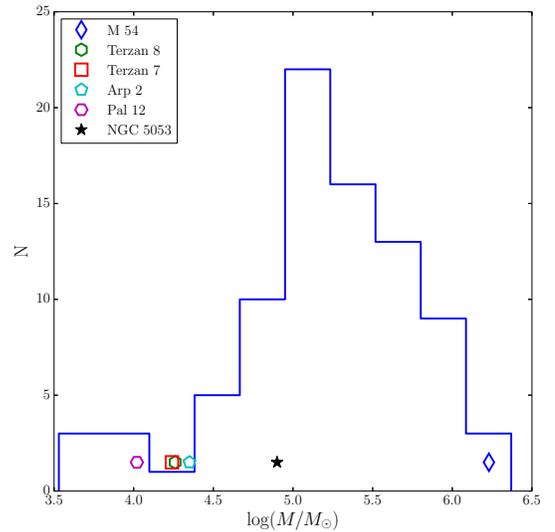}
\caption{Distribution of GC masses in the Milky Way. The locations of the 4 Sgr dSph GCs are marked in the distribution with
open points and NGC~5053 is marked with a filled star symbol. The distribution was taken from \cite{2005ApJS..161..304M}.\label{fig13} }
\end{figure}

\begin{figure*}
\includegraphics[width=\textwidth]{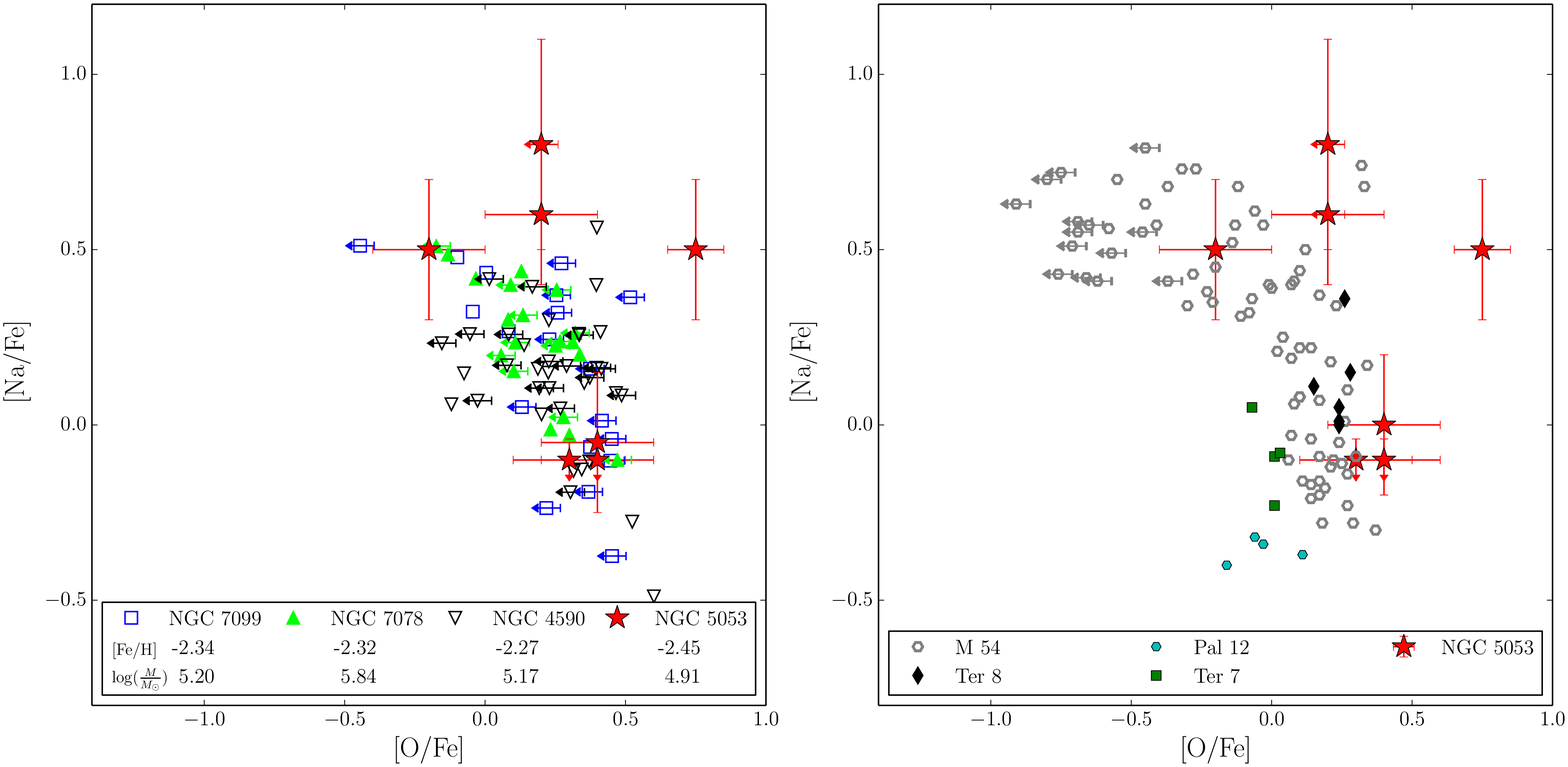}
\caption{Left panel: plot of the [Na/Fe] vs. [O/Fe] abundances in NGC~5053, NGC~7099, NGC~7078, NGC~4590. The [Fe/H] abundance and mass for each cluster
is given in the plot legend.
Right panel: plot of [Na/Fe] vs. [O/Fe] for NGC~5053 and other Sgr dSph candidate GCs. The abundances
for M54, Pal~12, Terzan~7, and Terzan~8 were taken from \cite{2010A&A...520A..95C}, \cite{2004AJ....127.1545C}, \cite{2007A&A...465..815S}, 
and \cite{2014A&A...561A..87C}, respectively. All of the abundances have been put on
our abundance scale and corrected for NLTE affects. \label{fig14} }
\end{figure*}

NGC~5053 would also be a relatively high-mass GC member of the Sgr dSph. 
Plotted in Figure 13 is the distribution of GC masses in the MW. The locations
of M54, Arp~2, Terzan~7, Terzan~8, Pal~12 and NGC~5053 are 
marked on the distribution. The MW GC distribution shown in this plot
results from the King model masses reported by \cite{2005ApJS..161..304M}.
The individual masses marked on the distribution were calculated based on their
respective absolute magnitudes reported in Harris (1996, Version 2010) and a 
mass-to-light ratio of $M/L = 2$. We used these calculated masses so that the
individual clusters would be on a consistent scale. It should be noted that
recent work by \cite{2012MNRAS.421..960S} and \cite{2014MNRAS.443.1425S}
report the mass of Terzan 8 to be slightly higher at $ \mathrm{log} (M/M_{\odot}) = $
4.5 and 4.93, respectively. The result by \cite{2014MNRAS.443.1425S} would give Terzan 8
a mass approximately equal to, or slightly larger, than NGC~5053.
This figure shows that NGC~5053 would be the second most massive GC associated
with the Sgr dSph, after M54.

\subsection{Na--O Anti-correlation}
As stated earlier, it was only possible to measure the [Na/Fe] and [O/Fe] 
abundances for seven of the stars in our sample. Despite this limited
sample, NGC~5053 still exhibits an abundance pattern in these species that is
consistent with the Na--O anti-correlation found in other clusters with larger
samples, such as M54. In the left-hand panel of Figure 14, we plot
the [Na/Fe] vs [O/Fe] abundances for NGC~5053 and clusters of similar metallicity (NGC 7099, NGC 7078, NGC 4590) from \cite{2009A&A...505..139C}.
This panel shows that the extent of the
Na--O anti-correlation present in NGC~5053 is consistent with that found in other
metal-poor MW GCs. In the right-hand panel of Figure 14 we plot the [Na/Fe] and [O/Fe] abundances for 
NGC~5053 and the Sgr dSph clusters M54, Pal~12, Terzan~7, and Terzan~8. 
Pal~12 is an additional cluster that is considered to be a likely candidate associated with the Sgr dSph 
by \cite{2010ApJ...718.1128L}. The abundances from other studies have been
placed on our assumed solar scale and the [Na/Fe] abundances have been
corrected for NLTE effects following the models in \cite{1999A&A...350..955G}.
From this panel, we see that Terzan~7, Terzan~8, and Pal~12 do not show an extended Na--O anti-correlation and
appear to be composed mostly, or entirely, of a single, first generation (FG) 
of stars. \cite{2014A&A...561A..87C} suggest that Terzan~8 is a GC composed 
of mostly FG stars, with only 1 star in their sample appearing to belong to the
SG. The size of our sample of NGC~5053 stars with Na and O abundances is 
comparable to the \cite{2014A&A...561A..87C} sample, with 7 and 6 respectively.
The NGC~5053 sample, however, shows signs of a more extended Na--O 
anti-correlation and a larger fraction of stars in the SG. 
This panel also shows that the apparent FG of stars in 
NGC~5053 has [Na/Fe] abundances consistent with those in Pal~12, Terzan~7, 
and Terzan~8.

With so few stars in the sample it is difficult to put constraints 
on how to divide the sample into the FG and SG stars.
Based on the [Na/Fe] distribution shown in Figure 10, an [Na/Fe] 
abundance of $\sim 0.4$ roughly divides the double peaked Gaussian distribution
into its two components, with an [Na/Fe] $< 0.4$ being the 
FG and [Na/Fe] $> 0.4$ being the SG of stars. Following this separation 
criterion, our sample would contain 3 FG stars and 8 SG
stars. The fractions of FG and SG stars in NGC~5053 are consistent with 
those found in other MW GCs \citep{2009A&A...505..117C}. In Figure 3,
we show the distribution of these stars on the cluster field;
the FG stars are marked with blue triangles pointing up, and the SG stars are marked 
with red triangles pointing down. Unfortunately, 
there were too few stars to explore the radial distribution 
of the different generations in the cluster. Out of the 5 GCs plotted 
in Figure 14, NGC~5053 is the second most massive, after M54. 
In this less massive Sgr dSph candidate GC sample, 
NGC~5053 is the only GC to show signatures of SG stars. 
While this does not put any constraints on the likelihood that NGC~5053 is a 
Sgr dSph cluster, it does illustrate how cluster mass affects and balances the 
presence of multiple generations.

A study by \cite{2011AJ....142..126S} found that NGC~5053 cluster members have an 
approximately unimodal CN band strength distribution, suggesting only FG stars are in the cluster.
While their result my seem to contradict what we find in NGC~5053, this is not the case. As noted by 
\cite{2011AJ....142..126S} and \cite{2010AJ....140.1119S}, 
their analysis of CN and CH molecular bands is limited by the [Fe/H] of the clusters. At low metallicities, such as those found in NGC~5053,
the variations in these molecular band strengths will become difficult to detect, causing clusters to appear to be chemically homogeneous in
light element abundances. This leads them to conclude that a low metallicity cluster showing a unimodal band strength distribution 
could still have chemical inhomogeneities and multiple populations in the cluster.
\\

\section{Conclusions}

NGC~5053 is possibly a very dynamically complex GC in our Galaxy. 
Not only has this cluster been shown to have significant tidal tails, but one 
study has also detected a tidal bridge with the neighboring 
cluster M53. The already rich dynamical environment of NGC~5053 is further
complicated by its possible association with the Sgr dSph. 
The goal of this study was to determine if NGC~5053 shows the typical chemical
signatures of MW GCs by collecting high quality 
spectra for a sample of RGB stars in the cluster. By using both EW measurements
and spectral synthesis, we have been able to determine 
the Fe, Ca, Ti, Ni, Ba, Na, and O abundances in NGC~5053.

The abundance results are summarized as follows:

\begin{enumerate}
 \item The cluster average [Fe/H] = -2.45 from 15 stars, with $\sigma =0.04$, making NGC~5053 one of the most
 metal-poor GCs in the Galaxy.
 \item From our target selection we determine the radial velocity of the cluster to be
 $v_{r} = 42.0 \pm 1.4$ $\mathrm{km \, s^{-1}}$.
 \item At this metallicity, the [Ca/Fe], [Ti/Fe], and [Ba/Fe] abundances are consistent with
 the abundances found for halo stars in the MW. Specifically, the $\alpha$-elements are enhanced
 and the Ba is depleted, relative to solar.
 \item The Na and O abundances exhibit star-to-star 
 variations within our sample. These variations 
 are larger than the typical uncertainty in their respective 
 measurements. The [Na/Fe] distribution is well fit 
 by two Gaussians with mean values separated by 0.8 dex. 
 The Na and O abundances exhibit a distribution that is 
 consistent with other GCs in the MW. From these abundances, NGC~5053
 appears to have both a FG and a SG of stars still present in the cluster. 
 The relative fraction of FG and SG stars in our sample for NGC~5053
 is consistent with other GCs in the MW. The extent of
 the Na-O anti-correlation in NGC~5053 is similar to what is seen in 
 other metal-poor MW GCs, specifically M15, M68, and M30.
\item Like the other Sgr dSph candidate GCs, NGC~5053 appears chemically similar
 (e.g., alpha-enhanced) to the MW field population at a similar 
 metallicity. Unlike MW GCs, however, the other Sgr dSph GC 
 candidates do not show an extended Na--O anti-correlation 
 or clear signs of multiple populations, apart from M54. 
 Out of the candidate GCs with abundances we pulled from the literature, 
 NGC~5053 is the most massive, after M54, and shows the signature 
 of the most extended Na-O anti-correlation.
 In the case of the other Sgr dSph candidate GCs, 
 this could be due to the small sample sizes available and the influence of
 their relatively low masses. 
 \item Based on these pieces of evidence we
 are unable to put further constraints on the possibility that NGC~5053 is
 a Sgr dSph cluster, so it will have to remain a maybe. 
 While the Na--O anti-correlation we see is similar to
 those seen in other MW GCs, and not in the low mass Sgr dSph GCs candidates, 
 this does not rule out its membership with the Sgr dSph. This could be the result of 
 NGC~5053 being massive enough to support multiple generations while the less massive Sgr dSph clusters are not.
 These results do suggest, however, that the possible complex dynamical 
 history of NGC~5053 has not affected its ability
 to produce the abundance trends we expect to see in GCs and the MW.
\end{enumerate}

This work is the first piece of a comprehensive study of NGC~5053 and its
neighbor M~53. Using WIYN, we have collected deep, wide-field
photometry of the area around the two clusters, as well as the area
between them, to fully charcterize the morphological properties of 
their outer regions and put further constraints on their tidal tails 
and the possible
tidal bridge. Additionally, we have collected Hydra data for M53 that
will allow us to explore its abundances in a similar way to this study.
We will pair these observational data with N-body simulations of the clusters
to create a more dynamically complete picture of the evolution
of these clusters and the origin of their tidal features.
\nocite{*}

\acknowledgments
We would like to thank to Anna Lisa Varri for the initial discussions 
that helped seed this project, Maria J. Cordero who assisted with the observations and data
reduction, and Jamie Overbeek who gave guidance on how to determine Ba abundances.
EV acknowledges support by grant NASA NNX13AF45G.
This material is based on work supported by the National Science Foundation
Graduate Research Fellowship Program under Grant No. DGE-1342962 to OB. Any 
opinion, findings, and conclusions or recommendations expressed in this material
are those of the author(s) and do not necessarily reflect the views of the 
National Science~Foundation.

\bibliography{boberg2015}

\begin{thebibliography}{}
\expandafter\ifx\csname natexlab\endcsname\relax\def\natexlab#1{#1}\fi

\bibitem[{{Alonso} {et~al.}(1999){Alonso}, {Arribas}, \&
  {Mart{\'{\i}}nez-Roger}}]{1999A&AS..140..261A}
{Alonso}, A., {Arribas}, S., \& {Mart{\'{\i}}nez-Roger}, C. 1999, \aaps, 140,
  261

\bibitem[{{Anders} \& {Grevesse}(1989)}]{1989GeCoA..53..197A}
{Anders}, E., \& {Grevesse}, N. 1989, \gca, 53, 197

\bibitem[{{Armandroff} {et~al.}(1992){Armandroff}, {Da Costa}, \&
  {Zinn}}]{1992AJ....104..164A}
{Armandroff}, T.~E., {Da Costa}, G.~S., \& {Zinn}, R. 1992, \aj, 104, 164

\bibitem[{{Carretta} {et~al.}(2009{\natexlab{a}}){Carretta}, {Bragaglia},
  {Gratton}, \& {Lucatello}}]{2009A&A...505..139C}
{Carretta}, E., {Bragaglia}, A., {Gratton}, R., \& {Lucatello}, S.
  2009{\natexlab{a}}, \aap, 505, 139

\bibitem[{{Carretta} {et~al.}(2010{\natexlab{a}}){Carretta}, {Bragaglia},
  {Gratton}, {Lucatello}, {Bellazzini}, \& {D'Orazi}}]{2010ApJ...712L..21C}
{Carretta}, E., {Bragaglia}, A., {Gratton}, R., {et~al.} 2010{\natexlab{a}},
  \apjl, 712, L21

\bibitem[{{Carretta} {et~al.}(2014){Carretta}, {Bragaglia}, {Gratton},
  {D'Orazi}, {Lucatello}, \& {Sollima}}]{2014A&A...561A..87C}
{Carretta}, E., {Bragaglia}, A., {Gratton}, R.~G., {et~al.} 2014, \aap, 561,
  A87

\bibitem[{{Carretta} {et~al.}(2010{\natexlab{b}}){Carretta}, {Bragaglia},
  {Gratton}, {Lucatello}, \& {Bellazzini}}]{2010A&A...520A..95C}
{Carretta}, E., {Bragaglia}, A., {Gratton}, R.~G., {Lucatello}, S., \&
  {Bellazzini}. 2010{\natexlab{b}}, \aap, 520, A95

\bibitem[{{Carretta} {et~al.}(2004){Carretta}, {Bragaglia}, {Gratton}, \&
  {Tosi}}]{2004A&A...422..951C}
{Carretta}, E., {Bragaglia}, A., {Gratton}, R.~G., \& {Tosi}, M. 2004, \aap,
  422, 951

\bibitem[{{Carretta} {et~al.}(2009{\natexlab{b}}){Carretta}, {Bragaglia},
  {Gratton}, {Lucatello}, {Catanzaro}, {Leone}, {Bellazzini}, {Claudi},
  {D'Orazi}, {Momany}, {Ortolani}, {Pancino}, {Piotto}, {Recio-Blanco}, \&
  {Sabbi}}]{2009A&A...505..117C}
{Carretta}, E., {Bragaglia}, A., {Gratton}, R.~G., {et~al.} 2009{\natexlab{b}},
  \aap, 505, 117

\bibitem[{{Chun} {et~al.}(2010){Chun}, {Kim}, {Sohn}, {Park}, {Han}, {Kim},
  {Lee}, {Lee}, {Lee}, \& {Sohn}}]{2010AJ....139..606C}
{Chun}, S.-H., {Kim}, J.-W., {Sohn}, S.~T., {et~al.} 2010, \aj, 139, 606

\bibitem[{{Cohen}(2004)}]{2004AJ....127.1545C}
{Cohen}, J.~G. 2004, \aj, 127, 1545

\bibitem[{{Cohen} \& {Kirby}(2012)}]{2012ApJ...760...86C}
{Cohen}, J.~G., \& {Kirby}, E.~N. 2012, \apj, 760, 86

\bibitem[{{Da Costa} \& {Armandroff}(1995)}]{1995AJ....109.2533D}
{Da Costa}, G.~S., \& {Armandroff}, T.~E. 1995, \aj, 109, 2533

\bibitem[{{de Mink} {et~al.}(2009){de Mink}, {Pols}, {Langer}, \&
  {Izzard}}]{2009A&A...507L...1D}
{de Mink}, S.~E., {Pols}, O.~R., {Langer}, N., \& {Izzard}, R.~G. 2009, \aap,
  507, L1

\bibitem[{{D'Ercole} {et~al.}(2012){D'Ercole}, {D'Antona}, {Carini},
  {Vesperini}, \& {Ventura}}]{2012MNRAS.423.1521D}
{D'Ercole}, A., {D'Antona}, F., {Carini}, R., {Vesperini}, E., \& {Ventura}, P.
  2012, \mnras, 423, 1521

\bibitem[{{D'Ercole} {et~al.}(2010){D'Ercole}, {D'Antona}, {Ventura},
  {Vesperini}, \& {McMillan}}]{2010MNRAS.407..854D}
{D'Ercole}, A., {D'Antona}, F., {Ventura}, P., {Vesperini}, E., \& {McMillan},
  S.~L.~W. 2010, \mnras, 407, 854

\bibitem[{{D'Ercole} {et~al.}(2008){D'Ercole}, {Vesperini}, {D'Antona},
  {McMillan}, \& {Recchi}}]{2008MNRAS.391..825D}
{D'Ercole}, A., {Vesperini}, E., {D'Antona}, F., {McMillan}, S.~L.~W., \&
  {Recchi}, S. 2008, \mnras, 391, 825

\bibitem[{{Geisler} {et~al.}(1995){Geisler}, {Piatti}, {Claria}, \&
  {Minniti}}]{1995AJ....109..605G}
{Geisler}, D., {Piatti}, A.~E., {Claria}, J.~J., \& {Minniti}, D. 1995, \aj,
  109, 605

\bibitem[{{Gratton} {et~al.}(2012){Gratton}, {Carretta}, \&
  {Bragaglia}}]{2012A&ARv..20...50G}
{Gratton}, R.~G., {Carretta}, E., \& {Bragaglia}, A. 2012, \aapr, 20, 50

\bibitem[{{Gratton} {et~al.}(1999){Gratton}, {Carretta}, {Eriksson}, \&
  {Gustafsson}}]{1999A&A...350..955G}
{Gratton}, R.~G., {Carretta}, E., {Eriksson}, K., \& {Gustafsson}, B. 1999,
  \aap, 350, 955

\bibitem[{{Gustafsson} {et~al.}(2008){Gustafsson}, {Edvardsson}, {Eriksson},
  {J{\o}rgensen}, {Nordlund}, \& {Plez}}]{2008A&A...486..951G}
{Gustafsson}, B., {Edvardsson}, B., {Eriksson}, K., {et~al.} 2008, \aap, 486,
  951

\bibitem[{{Harris}(1996)}]{1996AJ....112.1487H}
{Harris}, W.~E. 1996, \aj, 112, 1487

\bibitem[{{Ibata} {et~al.}(1994){Ibata}, {Gilmore}, \&
  {Irwin}}]{1994Natur.370..194I}
{Ibata}, R.~A., {Gilmore}, G., \& {Irwin}, M.~J. 1994, \nat, 370, 194

\bibitem[{{Johnson} {et~al.}(2008){Johnson}, {Pilachowski}, {Simmerer}, \&
  {Schwenk}}]{2008ApJ...681.1505J}
{Johnson}, C.~I., {Pilachowski}, C.~A., {Simmerer}, J., \& {Schwenk}, D. 2008,
  \apj, 681, 1505

\bibitem[{{Jordi} \& {Grebel}(2010)}]{2010A&A...522A..71J}
{Jordi}, K., \& {Grebel}, E.~K. 2010, \aap, 522, A71

\bibitem[{{Kimmig} {et~al.}(2014){Kimmig}, {Seth}, {Ivans}, {Strader},
  {Caldwell}, {Anderton}, \& {Gregersen}}]{2014arXiv1411.1763K}
{Kimmig}, B., {Seth}, A., {Ivans}, I.~I., {et~al.} 2014, ArXiv e-prints,
  arXiv:1411.1763

\bibitem[{{Lauchner} {et~al.}(2006){Lauchner}, {Powell}, \&
  {Wilhelm}}]{2006ApJ...651L..33L}
{Lauchner}, A., {Powell}, Jr., W.~L., \& {Wilhelm}, R. 2006, \apjl, 651, L33

\bibitem[{{Law} \& {Majewski}(2010)}]{2010ApJ...718.1128L}
{Law}, D.~R., \& {Majewski}, S.~R. 2010, \apj, 718, 1128

\bibitem[{{McLaughlin} \& {van der Marel}(2005)}]{2005ApJS..161..304M}
{McLaughlin}, D.~E., \& {van der Marel}, R.~P. 2005, \apjs, 161, 304

\bibitem[{{McWilliam}(1998)}]{1998AJ....115.1640M}
{McWilliam}, A. 1998, \aj, 115, 1640

\bibitem[{{Mottini} {et~al.}(2008){Mottini}, {Wallerstein}, \&
  {McWilliam}}]{2008AJ....136..614M}
{Mottini}, M., {Wallerstein}, G., \& {McWilliam}, A. 2008, \aj, 136, 614

\bibitem[{{Odenkirchen} {et~al.}(2001){Odenkirchen}, {Grebel}, {Rockosi},
  {Dehnen}, {Ibata}, {Rix}, {Stolte}, {Wolf}, {Anderson}, {Bahcall},
  {Brinkmann}, {Csabai}, {Hennessy}, {Hindsley}, {Ivezi{\'c}}, {Lupton},
  {Munn}, {Pier}, {Stoughton}, \& {York}}]{2001ApJ...548L.165O}
{Odenkirchen}, M., {Grebel}, E.~K., {Rockosi}, C.~M., {et~al.} 2001, \apjl,
  548, L165

\bibitem[{{Pilachowski} {et~al.}(1996){Pilachowski}, {Sneden}, \&
  {Kraft}}]{1996AJ....111.1689P}
{Pilachowski}, C.~A., {Sneden}, C., \& {Kraft}, R.~P. 1996, \aj, 111, 1689

\bibitem[{{Piotto} {et~al.}(2014){Piotto}, {Milone}, {Bedin}, {Anderson},
  {King}, {Marino}, {Nardiello}, {Aparicio}, {Barbuy}, {Bellini}, {Brown},
  {Cassisi}, {Cunial}, {Dalessandro}, {D'Antona}, {Ferraro}, {Hidalgo},
  {Lanzoni}, {Monelli}, {Ortolani}, {Renzini}, {Salaris}, {Sarajedini}, {van
  der Marel}, {Vesperini}, \& {Zoccali}}]{2014arXiv1410.4564P}
{Piotto}, G., {Milone}, A.~P., {Bedin}, L.~R., {et~al.} 2014, ArXiv e-prints,
  arXiv:1410.4564

\bibitem[{{Prantzos} \& {Charbonnel}(2006)}]{2006A&A...458..135P}
{Prantzos}, N., \& {Charbonnel}, C. 2006, \aap, 458, 135

\bibitem[{{Preston} {et~al.}(2006){Preston}, {Sneden}, {Thompson}, {Shectman},
  \& {Burley}}]{2006AJ....132...85P}
{Preston}, G.~W., {Sneden}, C., {Thompson}, I.~B., {Shectman}, S.~A., \&
  {Burley}, G.~S. 2006, \aj, 132, 85

\bibitem[{{Rieke} \& {Lebofsky}(1985)}]{1985ApJ...288..618R}
{Rieke}, G.~H., \& {Lebofsky}, M.~J. 1985, \apj, 288, 618

\bibitem[{{Salinas} {et~al.}(2012){Salinas}, {J{\'{\i}}lkov{\'a}}, {Carraro},
  {Catelan}, \& {Amigo}}]{2012MNRAS.421..960S}
{Salinas}, R., {J{\'{\i}}lkov{\'a}}, L., {Carraro}, G., {Catelan}, M., \&
  {Amigo}, P. 2012, \mnras, 421, 960

\bibitem[{{Sarajedini} \& {Milone}(1995)}]{1995AJ....109..269S}
{Sarajedini}, A., \& {Milone}, A.~A.~E. 1995, \aj, 109, 269

\bibitem[{{Sarajedini} {et~al.}(2007){Sarajedini}, {Bedin}, {Chaboyer},
  {Dotter}, {Siegel}, {Anderson}, {Aparicio}, {King}, {Majewski},
  {Mar{\'{\i}}n-Franch}, {Piotto}, {Reid}, \&
  {Rosenberg}}]{2007AJ....133.1658S}
{Sarajedini}, A., {Bedin}, L.~R., {Chaboyer}, B., {et~al.} 2007, \aj, 133, 1658

\bibitem[{{Sbordone} {et~al.}(2007){Sbordone}, {Bonifacio}, {Buonanno},
  {Marconi}, {Monaco}, \& {Zaggia}}]{2007A&A...465..815S}
{Sbordone}, L., {Bonifacio}, P., {Buonanno}, R., {et~al.} 2007, \aap, 465, 815

\bibitem[{{Shetrone} {et~al.}(2010){Shetrone}, {Martell}, {Wilkerson}, {Adams},
  {Siegel}, {Smith}, \& {Bond}}]{2010AJ....140.1119S}
{Shetrone}, M., {Martell}, S.~L., {Wilkerson}, R., {et~al.} 2010, \aj, 140,
  1119

\bibitem[{{Smolinski} {et~al.}(2011){Smolinski}, {Martell}, {Beers}, \&
  {Lee}}]{2011AJ....142..126S}
{Smolinski}, J.~P., {Martell}, S.~L., {Beers}, T.~C., \& {Lee}, Y.~S. 2011,
  \aj, 142, 126

\bibitem[{{Sneden}(1973)}]{1973ApJ...184..839S}
{Sneden}, C. 1973, \apj, 184, 839

\bibitem[{{Sobeck} {et~al.}(2011){Sobeck}, {Kraft}, {Sneden}, {Preston},
  {Cowan}, {Smith}, {Thompson}, {Shectman}, \& {Burley}}]{2011AJ....141..175S}
{Sobeck}, J.~S., {Kraft}, R.~P., {Sneden}, C., {et~al.} 2011, \aj, 141, 175

\bibitem[{{Sollima} {et~al.}(2014){Sollima}, {Carretta}, {D'Orazi}, {Gratton},
  {Bragaglia}, \& {Lucatello}}]{2014MNRAS.443.1425S}
{Sollima}, A., {Carretta}, E., {D'Orazi}, V., {et~al.} 2014, \mnras, 443, 1425

\bibitem[{{Suntzeff} {et~al.}(1988){Suntzeff}, {Kraft}, \&
  {Kinman}}]{1988AJ.....95...91S}
{Suntzeff}, N.~B., {Kraft}, R.~P., \& {Kinman}, T.~D. 1988, \aj, 95, 91

\bibitem[{{Venn} {et~al.}(2004){Venn}, {Irwin}, {Shetrone}, {Tout}, {Hill}, \&
  {Tolstoy}}]{2004AJ....128.1177V}
{Venn}, K.~A., {Irwin}, M., {Shetrone}, M.~D., {et~al.} 2004, \aj, 128, 1177

\bibitem[{{Ventura} \& {D'Antona}(2008)}]{2008A&A...479..805V}
{Ventura}, P., \& {D'Antona}, F. 2008, \aap, 479, 805

\bibitem[{{Ventura} {et~al.}(2001){Ventura}, {D'Antona}, {Mazzitelli}, \&
  {Gratton}}]{2001ApJ...550L..65V}
{Ventura}, P., {D'Antona}, F., {Mazzitelli}, I., \& {Gratton}, R. 2001, \apjl,
  550, L65

\bibitem[{{Wallace} {et~al.}(2011){Wallace}, {Hinkle}, {Livingston}, \&
  {Davis}}]{2011ApJS..195....6W}
{Wallace}, L., {Hinkle}, K.~H., {Livingston}, W.~C., \& {Davis}, S.~P. 2011,
  \apjs, 195, 6

\bibitem[{{Zinn}(1985)}]{1985ApJ...293..424Z}
{Zinn}, R. 1985, \apj, 293, 424

\end{thebibliography}

\end{document}